\documentclass[12pt,a4paper]{article}

\usepackage{geometry}
\usepackage[symbol]{footmisc}
\RequirePackage{amsmath}
\RequirePackage{mathrsfs}
\RequirePackage{amsfonts}
\RequirePackage{dsfont}
\RequirePackage{braket}
\RequirePackage{tabularx}
\usepackage{multirow}
\RequirePackage{nicefrac}
\RequirePackage{graphicx}
\RequirePackage{booktabs}
\RequirePackage{colortbl}
\RequirePackage{xcolor}
\RequirePackage[symbol]{footmisc}
\usepackage{mathtools}
\usepackage{comment}
\usepackage{blkarray}
\usepackage{stackengine}
\usepackage{float}
\usepackage{rotating}
\usepackage{hhline}
\usepackage[section]{placeins}
\def\AEF{A.E. Faraggi}

\def\EJP#1#2#3{{\it Eur.\ Phys.\ Jour.}\/ {\bf C#1} (#2) #3}

\def\JHEP#1#2#3{{\it JHEP}\/ {\bf #1} (#2) #3}
\def\NPB#1#2#3{{\it Nucl.\ Phys.}\/ {\bf B#1} (#2) #3}
\def\PLB#1#2#3{{\it Phys.\ Lett.}\/ {\bf B#1} (#2) #3}
\def\PRD#1#2#3{{\it Phys.\ Rev.}\/ {\bf D#1} (#2) #3}
\def\PRL#1#2#3{{\it Phys.\ Rev.\ Lett.}\/ {\bf #1} (#2) #3}

\def\etal{{\it et al\/}}

\def\beq{\begin{equation}}
\def\eeq{\end{equation}}
\def\beqn{\begin{eqnarray}}
\def\eeqn{\end{eqnarray}}
\def\ds{{$\tilde S$}}
\def\unahe{{${\overline{\rm NAHE}}$}}

\newcommand{\CC}[2]{C{#1\atopwithdelims[]#2}}

\newcommand{\ba}{\begin{eqnarray}}
\newcommand{\ea}{\end{eqnarray}}
\DeclareRobustCommand{\sqbinom}{\genfrac[]{0pt}{}}

\numberwithin{equation}{section}

\begin{document}
\begin{titlepage}
\samepage{
\setcounter{page}{1}
\rightline{}
\rightline{November 2020}

\vfill
\begin{center}
  {\Large \bf{ Classification of \\ \medskip
      Non-Supersymmetric \\ \bigskip
      Pati-Salam Heterotic String Models 
      }}

\vspace{1cm}
\vfill

{\large Alon E. Faraggi\footnote{E-mail address: alon.faraggi@liv.ac.uk}, 
         Viktor G. Matyas\footnote{E-mail address: viktor.matyas@liv.ac.uk}
 and Benjamin Percival\footnote{E-mail address: benjamin.percival@liv.ac.uk} 
}
\\

\vspace{1cm}

{\it Dept.\ of Mathematical Sciences, University of Liverpool, Liverpool
L69 7ZL, UK\\}

\vspace{.025in}
\end{center}

\vfill
\begin{abstract}
\noindent

We extend the classification of fermionic $\mathbb{Z}_2\times\mathbb{Z}_2$ 
heterotic string orbifolds to
non--supersymmetric Pati--Salam (PS) models in two classes of vacua, 
that we dub \ds--models and $S$--models. 
The first correspond to compactifications of a tachyonic ten--dimensional
vacuum, whereas the second 
correspond to compactifications of the ten--dimensional tachyon--free 
$SO(16)\times SO(16)$ heterotic string. In both cases we develop a systematic
method to 
extract tachyon--free four--dimensional models. We show that tachyon--free
configurations 
arise with probability $\sim0.002$ and $\sim0.01$ in the first and second
case, respectively. 
We adapt the `fertility methodology' that facilitates the extraction of
phenomenological 
models. We show that Pati--Salam \ds--models do not contain heavy Higgs
scalar 
representations that are required to break the PS symmetry to the Standard
Model and are therefore
not phenomenologically viable. Hence, we argue that in \ds--models the
$SO(10)$ GUT symmetry
must be broken at the string scale to the Standard--like Model subgroup. 
We extract tachyon--free three generation models in both cases 
that contain an equal number of massless bosonic and fermionic degrees of
freedom, {\it i.e.}
with $a_{00}=N_b^0-N_f^0=0$, and analyse their one--loop partition function.

\end{abstract}

\smallskip}

\end{titlepage}

\section{Introduction}\label{intro}
The Standard Model of particle physics provides viable parameterisation
of most observable sub--atomic data to date. Nevertheless, it is clear that
the Standard Model is merely an effective field theory description, albeit
one that may be effective up to the Grand Unified Theory (GUT),
or Planck, scales. The first evidence for the need to go beyond the Standard
Model stems from the observation of non--vanishing neutrino masses, which
requires the augmentation of the Standard Model spectrum with right--handed
neutrinos, or the introduction of nonrenormalisable interactions,
that are suppressed by some cutoff scale, at which the Standard Model
ceases to provide viable parameterisation.
Ultimately, the Standard Model is formulated using the framework of
point quantum field theory, which is fundamentally incompatible
with gravity.

An early construction to accommodate non--trivial neutrino masses
was the Pati--Salam model \cite{ps} that sought to introduce a symmetry
between quarks and leptons. This rightly celebrated insight was vindicated
by experiments \cite{nmevidence}. It paved the road for the so--called
Grand Unified Theories (GUTs) that aim to unify all the sub--atomic
gauge interactions into a simple or semi--simple gauge group \cite{raby}.
Experimental constraints stemming from proton longevity and suppression
of left--handed neutrino masses mandate that the unification scale
is much higher than the electroweak scale, which is currently being
actively probed by experiments. The GUT scale is in fact an order of
magnitude or two below the Planck scale, the scale at which the gauge
and gravitational interactions are of comparable strength. 

It is therefore sensible to construct GUT models in a framework
that incorporates gravity into the fold. 
Building quantum field theory models at a scale which is one or two orders
of magnitude
below the Planck scale is fraught with uncertainties as quantum gravity
effects may become dominant. String theory provides a framework that enables
the construction of particle physics models within a perturbatively
consistent theory of quantum gravity \cite{hetecan}. 
String theory gives rise to a large number of vacuum solutions that may
\textit{a priori} be relevant for the particle physics data.
The way forward is to explore the properties of classes of string vacua
that share some common characteristics and to develop the tools to discern them
from other classes. While the majority of models studied to date possess
$N=1$ spacetime supersymmetry, 
non--supersymmetric vacua that correspond to compactifications
of the tachyon--free ten--dimensional $SO(16)\times SO(16)$ heterotic string
have been of interest as well \cite{dh, itoyama, nonSUSY, aafs, ADM, FR}. 
It is well known that in addition
to the tachyon--free vacua, string theory gives rise to tachyonic
vacua in ten dimensions \cite{dh,kltclas, gv}. 
Recently, it was argued
that these ten--dimensional
vacua may serve as good starting points for the construction of
phenomenological models, provided that the tachyonic states are projected
out from the physical spectrum in the four--dimensional models,
and may offer some novel perspectives on some of the outstanding
issues in string phenomenology \cite{spwsp, stable, kaidi}. A three
generation standard--like model in this class was presented in
\cite{stable}, and it was argued that all the moduli are projected out in
the model and that it may therefore be stable \cite{stable}.

The free fermionic formulation \cite{fff} of the heterotic string gave rise to a
large space of phenomenological string vacua, which correspond to
compactifications of the ten--dimensional heterotic string
on $\mathbb{Z}_2\times\mathbb{Z}_2$ toroidal orbifolds. The initial constructions
corresponded to isolated examples with 
$SU(5)\times U(1)$ (FSU5) \cite{fsu5},
$SO(6)\times SO(4)$ (PS) \cite{alr},
$SU(3)\times SU(2)\times U(1)^2$ (SLM) \cite{fny,slm} and
$SU(3)\times U(1)\times SU(2)^2$ (LRS) \cite{lrs},
unbroken subgroups of $SO(10)$, with the
GUT embedding of the electroweak hypercharge, producing
the canonical prediction for the Weinberg angle
$\sin^2\theta_W=3/8$ at $M_{\rm String}$. 
Systematic methods
to classify large numbers of free fermionic models were developed
over the past two decades
\cite{gkr, fknr, fkr, acfkr, su62, frs, slmclass, lrsclass, ferlrs}.
In ref. \cite{so10tclass} the classification program was
extended to tachyon--free models with unbroken $SO(10)$ subgroup
that may be regarded as compactifications of a tachyonic ten--dimensional vacuum.
Numerous models with
equal number of massless bosons and fermions were found and
an analysis of the one loop vacuum amplitude was presented.
Of particular interest was the observation of the misaligned
supersymmetry phenomena \cite{missusy}, which guarantees the finiteness of the
string amplitude in the absence of spacetime supersymmetry.

In this paper we extend the free fermionic classification program to
non--supersymmetric models with PS gauge symmetry. We develop
the classification method both for compactifications of the
$SO(16)\times SO(16)$ heterotic string as well as for compactifications
of the tachyonic ten--dimensional vacuum. Preliminary development of the
methodology in the first case was carried out in ref. \cite{FR}, but
only with respect to simultaneous shifts of tori, whereas here we
include independent shifts with respect to the six internal circles. In the
fermionic language this entails enlarging the basis vectors that span
the models to include shifts with respect to circles, rather than tori. 

We show that the \ds--models cannot in fact give rise to viable models, 
due to the absence from the spectrum of the states required to break the 
PS gauge group to the Standard Model. We remark that this result
is similar to the absence of viable supersymmetric models with 
$SO(10)\rightarrow SU(4)\times SU(2)\times U(1)$ \cite{su421}. 

A particular focus of our analysis here is the adaptation of the
`fertility methodology' to the class of non--supersymmetric PS models. 
These methods were applied in the case of the SLM \cite{slmclass} and LRS 
\cite{ferlrs} models and is very efficient in extracting phenomenologically
viable vacua. 
The new element of the fertility methodology in the non--supersymmetric models
involves the extraction of tachyon--free models, which occur with low frequency
in the total space of vacua. To contrast the cases of the \ds--models with that 
of the $S$--models, we apply a parallel set of phenomenological criteria, despite the
fact that they do not hold in the case of the \ds--models due to the absence of 
Heavy Higgs scalar representations. In addition we analyse the partition function 
of the models. A particular interest is in models with equal number of massless bosons and 
fermions. We present exemplary three generation \ds-- and $S$--models that satisfy this 
criteria. 

Our paper is organised as follows: in Section \ref{10D} we discuss the general structure of 
\ds-- {\it vs} $S$--models. In Section \ref{StPS} we elaborate on the 
analysis of the massless spectrum in PS \ds--models, whereas in 
Section \ref{FertMeth} we discuss the fertility methodology in \ds--models. 
The corresponding analysis in the $S$--models is discussed in Section
\ref{PSNormS}.
In Section \ref{PF} we discuss the analysis of the partition functions. 
Section \ref{results} presents the results of the analysis and Section \ref{conclude} concludes 
our paper.

\section{ $S$ {\it vs} $\tilde{S}$--Models
}\label{10D}


We will explore four--dimensional
non-supersymmetric Pati--Salam models via two distinct
routes. Firstly, we will explore those PS models descending from the
tachyonic ten--dimensional heterotic string which utilize the vector $\tilde{S}$.
We will refer to these models as (PS) $\tilde{S}$-models.
Secondly, we will explore models derived from the $SO(16)\times SO(16)$
heterotic string which contain the SUSY--generating basis vector $S$ and
thus we will refer to these as $S$--models. 

String models in the free fermionic formulation are defined
in terms of boundary condition basis vectors and one--loop Generalised
GSO (GGSO) phases \cite{fff}. 
The $SO(16)\times SO(16)$ and $E_8\times E_8$
heterotic--models in ten dimensions are
specified in terms of a common set of basis vectors 
\ba
v_1={\mathds{1}}&=&\{\psi^\mu,\
\chi^{1,\dots,6} \ | \ \overline{\eta}^{1,2,3},
\overline{\psi}^{1,\dots,5},\overline{\phi}^{1,\dots,8}\},\nonumber\\
v_{2}=z_1&=&\{\overline{\psi}^{1,\dots,5},
              \overline{\eta}^{1,2,3} \},\nonumber\\
v_{3}=z_2&=&\{\overline{\phi}^{1,\dots,8}\},
\label{tendbasisvectors}
\ea 
where we employed the common notation used in the
free fermionic models
\cite{fsu5, slm, alr, lrs, gkr, acfkr, su62, frs, 
slmclass, lrsclass, fknr, fkr}. 
The spacetime supersymmetry generator arises from the combination 
\beq
S={\mathds{1}}+z_1+z_2 = \{{\psi^\mu},\chi^{1,\dots,6}\}. 
\label{tendsvector}
\eeq
The choice of GGSO phase $\CC{z_1}{z_2}=\pm 1$ selects between 
the $E_8\times E_8$ or $SO(16)\times SO(16)$ heterotic strings 
in ten dimensions.
Eq. (\ref{tendsvector}) implies that
the breaking of spacetime supersymmetry in ten dimensions
is correlated with the breaking pattern 
$E_8\times E_8\rightarrow SO(16)\times SO(16)$.
The would--be tachyons in these models arise from the Neveu-Schwarz (NS)
sector, by acting on the right--moving vacuum with a fermionic oscillator
\beq
\ket{0}_L\otimes {\bar\phi}^a\ket{0}_R,
\label{untwistedtach}
\eeq
where $a=1,\cdots, 16$.
The GSO projection induced by the $S$--vector projects 
out these NS tachyons,
producing tachyon--free models in both cases.

In lower dimensions eq.
(\ref{tendsvector}) does not hold and 
the two breakings are not correlated.
In lower dimensions
the projection of the $N=1$ supersymmetry generator
from the $S$--sector
is induced by an alternative phase. 
In this manner non--supersymmetric vacua are constructed.
In the orbifold representation, these free
fermion models 
correspond to compactifications of the ten--dimensional
$SO(16)\times SO(16)$ heterotic string, where the detailed dictionary
of ref. \cite{panosdictionary}, can be used to translate the free
fermion data to the toroidal orbifold data.
Tachyon producing sectors proliferate in the lower dimensional models
\cite{aafs}, and can be projected out for special choices of the GGSO
projection coefficients.

As discussed in refs. \cite{spwsp, stable, so10tclass}, the 
ten--dimensional tachyonic vacua in the free fermionic formulation are obtained
by removing the $S$--vector from the construction, or by augmenting it
with four periodic right--moving worldsheet fermions. 
These ten--dimensional configurations do contain the NS tachyons
appearing in eq. (\ref{untwistedtach}) or a subset of them, and are
connected by interpolations or orbifolds
along the lines of ref. \cite{gv, interpol}.
Similarly, 
the construction of the four--dimensional free fermion
models that descend from the ten--dimensional tachyonic vacua amounts to removing the vector $S$
from the set of basis vectors that are used to build the models.
Alternatively, the $S$--vector can be augmented with a set of four
right--moving periodic fermions \cite{spwsp, stable, so10tclass}.
A convenient choice is given by 
\beq
{\tilde S} = \{\psi^{1,2}, 
                \chi^{1,2},
                \chi^{3,4},
                \chi^{5,6} \ \vert \ {\bar\phi}^{3, ...,~6} \}
\label{newS}
\eeq
In this case there are no massless gravitinos, and the untwisted 
tachyonic states 
\beq
|0\rangle_L\otimes {\bar\phi}^{3, ...,~6}|0\rangle_R 
\label{stildetachstates}
\eeq
are invariant under the ${\tilde S}$--vector projection, and 
therefore are the NS tachyons that descend from the ten--dimensional vacuum.
We denote the general map, which is induced by the exchange
\beq
S\longleftrightarrow {\tilde S},
\label{stildemap}
\eeq
as the ${\tilde S}$--map. 
This map was used in the
\unahe--based model of ref. \cite{stable} and in the classification
of the $SO(10)$ GUT models in ref. \cite{so10tclass}.
As for the $S$--based models that correspond to compactifications
of the $SO(16)\times SO(16)$ heterotic string, the tachyonic
states in the \ds--models are projected out for discrete choices
of the GGSO projection coefficients.
Thus, these models should be
taken on equal footing as the $S$--models. However, the characteristics
of the spectra in the two cases are different. Our aim here is to explore
these different characteristics. 

\section{Pati--Salam $\tilde{S}$-Models}
\label{StPS}
Let us turn our attention to the classification set-up for the Pati--Salam $\tilde{S}$-models which descend from the 10D tachyonic heterotic string. We can build these models by first defining a set of 12 basis vectors that generate $SO(10)$ GUT $\tilde{S}$-models which were used in the classification performed in \cite{so10tclass}
\begin{align}\label{basisSt}
\mathds{1}&=\{\psi^\mu,\
\chi^{1,\dots,6},y^{1,\dots,6}, w^{1,\dots,6}\ | \ \overline{y}^{1,\dots,6},\overline{w}^{1,\dots,6},
\overline{\psi}^{1,\dots,5},\overline{\eta}^{1,2,3},\overline{\phi}^{1,\dots,8}\},\nonumber\\
\tilde{S}&=\{{\psi^\mu},\chi^{1,\dots,6} \ | \ \overline{\phi}^{3,4,5,6}\},\nonumber\\
{e_i}&=\{y^{i},w^{i}\; | \; \overline{y}^{i},\overline{w}^{i}\}, 
\ \ \ i=1,...,6
\nonumber\\
{b_1}&=\{\psi^\mu,\chi^{12},y^{34},y^{56}\; | \; \overline{y}^{34},
\overline{y}^{56},\overline{\eta}^1,\overline{\psi}^{1,\dots,5}\},\\
{b_2}&=\{\psi^\mu,\chi^{34},y^{12},y^{56}\; | \; \overline{y}^{12},
\overline{y}^{56},\overline{\eta}^2,\overline{\psi}^{1,\dots,5}\},\nonumber\\
{b_3}&=\{\psi^\mu,\chi^{56},y^{12},y^{34}\; | \; \overline{y}^{12},
\overline{y}^{34},\overline{\eta}^3,\overline{\psi}^{1,\dots,5}\},\nonumber\\
z_1&=\{\overline{\phi}^{1,\dots,4}\},\nonumber
\nonumber
\end{align}
where we note that the $\{ y^i,w^i \ || \ \bar{y}^i,\bar{w}^i\}$ are fermionised coordinates of the internal toroidal $\Gamma_{6,6}$ lattice such that $i\partial X^i_L =y^iw^i $ and the $e_i$'s allow for all symmetric shifts of the six internal circles.
Furthermore, we note the existence of a vector combination
\beq \label{z2}
z_2=\mathds{1}+\sum_{i=1}^6 e_i +\sum_{k=1}^3 b_k+z_1=\{\bar{\phi}^{5,6,7,8}\}
\eeq 
in these models, which is typically its own basis vector
in the supersymmetric classifications of \cite{fknr,acfkr,frs,slmclass,lrsclass,ferlrs}. 

In order to extend this structure to only those models realising the $SO(6)\times SO(4)$ PS subgroup of $SO(10)$ we supplement this basis with the vector:
\beq \label{alpha}
\alpha=\{\bar{\psi}^{4,5}, \bar{\phi}^{1,2}\}.
\eeq 
Having fixed our basis we can now specify a model by fixing a set of GGSO phases $\CC{v_i}{v_j}$, which for our Pati--Salam basis here involves 78 free phases with all others specified by modular invariance. Hence, the full space of models is of size $2^{78}\sim 3\times 10^{23}$ models. This is considerably larger than the space of supersymmetric Pati--Salam models classified in \cite{acfkr} where supersymmetry constraints reduce the parameter space. 

With the basis and GGSO phases fixed we can then construct the modular invariant 
Hilbert space $\mathcal{H}$ of states $\ket{S_\xi}$ for a model through the one--loop GGSO projections:
\begin{equation}
    \mathcal{H}=\bigoplus_{\xi\in\Xi}\prod^{k}_{i=1}
\left\{ e^{i\pi v_i\cdot F_{\xi}}\ket{S_\xi}=\delta_{\xi}
\CC{\xi}{v_i}^*\ket{S_\xi}\right\}
\end{equation}
where the $\xi$ are sectors formed as a linear combination of the basis vectors which span an additive group $\Xi$, $F_\xi$ is the fermion number operator and $\delta_\xi=1,-1$
is the spin-statistics index. 

The sectors in the model can be characterised according to the left and
right moving vacuum separately
\begin{align}
M_L^2&=-\frac{1}{2}+\frac{\xi_L \cdot\xi_L}{8}+N_L\\
M_R^2 &=-1+\frac{\xi_R \cdot\xi_R}{8}+N_R \nonumber
\end{align}
where $N_L$ and $N_R$ are sums over left and right moving oscillators, 
respectively. Physical states must then additionally satisfy the
Virasoro matching condition, $M_L^2=M_R^2$. States not satisfying this condition
are off--shell. 

It will be useful to distinguish between those features of our models depending on the presence of the PS breaking vector (\ref{alpha}) and those features present at the level of $SO(10)$ defined by the basis set (\ref{basisSt}) and explored in detail in our recent work \cite{so10tclass}. This distinction will be crucial in how we define our methodology for efficiently scanning the space of PS $\tilde{S}$--models using the so--called `fertility' methodology. 
\subsection{Gauge Group and Enhancements}
The untwisted sector gauge vector
bosons for our PS $\tilde{S}$--models give rise to the gauge group 
\beq 
SO(6)\times SO(4) \times  U(1)_1\times U(1)_2\times U(1)_3\times SO(4)^4.
\eeq 

The gauge group of a model may be enhanced by additional gauge bosons. The $SO(10)$ enhancements arise from $\psi^\mu \{\bar{\lambda}^i\}\ket{z_1},\psi^\mu \{\bar{\lambda}^i\}\ket{z_2}$ and $\psi^\mu \ket{z_1+z_2}$ sectors, where $\bar{\lambda}^i$ are right moving oscillators. These were discussed in ref. \cite{so10tclass} and so we will focus our analysis on enhancements at the PS level arising from:
\beq 
G_{\text{PS}}=
\begin{Bmatrix}
\psi^\mu\ket{\alpha}_L \otimes \{\bar{\lambda^i}\}\ket{\alpha}_R, \ \ \ \ \psi^\mu\ket{z_1+\alpha}_L \otimes \{\bar{\lambda^i}\}\ket{z_1+\alpha}_R \\
\psi^\mu \ket{z_2+\alpha}_L \otimes \ket{z_2+\alpha}_R, \\ \psi^\mu \ket{z_1+z_2+\alpha}_L \otimes \ket{z_1+z_2+\alpha}_R\\
\psi^\mu \ket{\tilde{S}+\tilde{x}+z_1+\alpha}_L \otimes \ket{\tilde{S}+\tilde{x}+z_1+\alpha}_R, \\
\psi^\mu \ket{\tilde{S}+\tilde{x}+z_1+z_2+\alpha}_L \otimes \ket{\tilde{S}+\tilde{x}+z_1+z_2+\alpha}_R
\end{Bmatrix}
\eeq
where we have made use of the important vector combination:
\beq \label{xtilde}
\tilde{x}=b_1+b_2+b_3=\{ \psi^\mu, \chi^{1,2,3,4,5,6} \ \Big| \ \bar{\psi}^{1,...,5},\bar{\eta}^{1,2,3}\}
\eeq  
the properties of which are discussed in some detail in ref. \cite{so10tclass}.

These enhancements are all `mixed' enhancements in the sense that they affect both the observable and hidden gauge group factors. The presence of these gauge bosons is model--dependent since it depends on
the survival of the sectors under the GGSO projections. 
It turns out that the survival of $\psi^\mu\ket{\alpha}_L \otimes \{\bar{\psi}^{1,2,3},\bar{\eta}^{1,2,3}\}\ket{\alpha}_R$ and $\psi^\mu\ket{z_1+\alpha}_L \otimes \{\bar{\psi}^{1,2,3},\bar{\eta}^{1,2,3}\}\ket{z_1+\alpha}_R $ is correlated exactly with the survival of the tachyonic sectors $\ket{\alpha}$ and $\ket{z_1+\alpha}$. Hence, the absence of tachyons necessitates the projection of these mixed gauge enhancements. We note that this is similar to the result for mixed $SO(10)$ enhancements 
$\psi^\mu\{\bar{\psi}^a\}\ket{z_1},\psi^\mu\{\bar{\psi}^a\} \ket{z_2},\ a=1,...,5 $, which must be projected for models free from $\ket{z_1}$ and $\ket{z_2}$ tachyonic sectors. However, the other right--moving oscillator cases for these enhancement sectors have to be checked carefully as they also can affect the observable gauge group. 
The analysis of the level--matched tachyonic
sectors in our PS case is presented in the following section. 
\subsection{Analysis of Tachyonic Sectors }\label{tsa}
Due to the absence of supersymmetry the analysis of whether on--shell 
tachyons arise in the spectrum of
our models is of utmost importance. Since the details of how to project the tachyonic sectors within
the $SO(10)$ construction are shown in \cite{so10tclass}, we will focus on the tachyonic sectors at
the PS level, \textit{i.e.} those arising due to the inclusion of the 
$\alpha$--vector. However, there
is one amendment to the projection conditions in Section 4 of ref. \cite{so10tclass} required, which
is that we must account for the fact that the $\alpha$--vector 
can be used as a projector for many of the
tachyonic sectors at the $SO(10)$--level. In particular, it can project any of the tachyonic sectors
except for those involving $z_1$.  This is an important detail when we turn to the fertility analysis
in Section \ref{FertMeth}.

Turning our attention to the tachyonic sectors introduced at the PS level, we note that on--shell tachyonic sectors will arise when 
\beq \label{onshelltach}
M_L^2=M_R^2<0,
\eeq 
which corresponds to left and right products of $\xi_L \cdot  \xi_L=0,1,2,3$ and $\xi_R \cdot  \xi_R=0,1,2,3,4,5,6,7$. The presence of such tachyonic sectors in the physical spectrum indicates the instability of the string vacuum with respect to the background on which the theory is defined.  

We can deduce that there are 84 PS tachyonic sectors for us to consider once the $SO(10)$ tachyons are projected. These sectors are displayed in Table \ref{StTachSecs}.  
\begin{center} 
\begin{table}[!ht]
\centering
\begin{tabular}{|c|c|}
\hline
Mass Level & Sectors \\
\hline
$(-1/2,-1/2)$& \ $\alpha,\ \ z_1+\alpha$\\ 
\hline
$(-3/8,-3/8)$& \ $e_i+\alpha,\ \ e_i+z_1+\alpha$\\ 
\hline
$(-1/4,-1/4)$& \ $e_i+e_j+\alpha,\ \ e_i+e_j+z_1+\alpha$\\ 
\hline
$(-1/8,-1/8)$& \ $e_i+e_j+e_k+\alpha,\ \  e_i+e_j+e_k+z_1+\alpha$\\ 
\hline
\end{tabular}
\caption{\label{StTachSecs} \emph{Level-matched PS tachyonic sectors and their
    mass level, where $i\neq j \neq k=1,...,6$}.}
\end{table}
\end{center}
In Table \ref{StTachConds} the conditions on the projection of these tachyonic sectors are delineated. We only show the cases when $i=1, \ j=2$ and $k=3$ as the other combinations are similar.\\
\newcommand\xrowht[2][0]{\addstackgap[.5\dimexpr#2\relax]{\vphantom{#1}}}
\begin{table}[!ht] 
\centering
\small
\setlength{\tabcolsep}{1.5pt}
\begin{tabular}{|c|l|}
\hline 
\textbf{Tachyonic Sector}&\textbf{Survival conditions}\\
\hline \xrowht[()]{10pt}
$\alpha$&$\CC{\alpha}{e_1}=\CC{\alpha}{e_2}=\CC{\alpha}{e_3}=\CC{\alpha}{e_4}=\CC{\alpha}{e_5}=\CC{\alpha}{e_6}=+1$\\
\xrowht[()]{10pt}
&$\CC{\alpha}{z_2}=\CC{\alpha}{z_1+\alpha+\tilde{x}}=+1$ \\ [0.75ex] 
\hline
\xrowht[()]{10pt}
$z_1+\alpha$&$\CC{z_1+\alpha}{e_1}=\CC{z_1+\alpha}{e_2}=\CC{z_1+\alpha}{e_3}=\CC{z_1+\alpha}{e_4}=\CC{z_1+\alpha}{e_5}=\CC{z_1+\alpha}{e_6}=+1$\\ 
\xrowht[()]{10pt}
&$\CC{z_1+\alpha}{z_2}=\CC{z_1+\alpha}{\alpha+\tilde{x}}=+1$\\
\hline
\xrowht[()]{10pt}
$e_1+\alpha $ &$\CC{e_1+\alpha}{e_2}=\CC{e_1+\alpha}{e_3}=\CC{e_1+\alpha}{e_4}=\CC{e_1+\alpha}{e_5}=\CC{e_1+\alpha}{e_6}=+1$\\ 
\xrowht[()]{10pt}
&$\CC{e_1+\alpha}{b_1+z_1+\alpha}=\CC{e_1+\alpha}{\tilde{x}+z_1+\alpha}=\CC{e_1+\alpha}{z_2}=+1$\\
\hline
\xrowht[()]{10pt}
$e_1+z_1+\alpha $ &$\CC{e_1+z_1+\alpha}{e_2}=\CC{e_1+z_1+\alpha}{e_3}=\CC{e_1+z_1+\alpha}{e_4}=\CC{e_1+z_1+\alpha}{e_5}=\CC{e_1+z_1+\alpha}{e_6}=+1$\\ 
\xrowht[()]{10pt}
&$\CC{e_1+z_1+\alpha}{b_1+\alpha}=\CC{e_1+z_1+\alpha}{\tilde{x}+\alpha}=\CC{e_1+z_1+\alpha}{z_2}=+1$\\
\hline
\xrowht[()]{10pt}
$e_1+e_2+\alpha $ &$\CC{e_1+e_2+\alpha}{e_3}=\CC{e_1+e_2+\alpha}{e_4}=\CC{e_1+e_2+\alpha}{e_5}=\CC{e_1+e_2+\alpha}{e_6}=+1$\\ 
\xrowht[()]{10pt}
&$\CC{e_1+e_2+\alpha}{b_1+z_1+\alpha}=\CC{e_1+e_2+\alpha}{\tilde{x}+z_1+\alpha}=\CC{e_1+e_2+\alpha}{z_2}=+1$\\
\hline
\xrowht[()]{10pt}
$e_1+e_2+z_1+\alpha $ &$\CC{e_1+e_2+z_1+\alpha}{e_3}=\CC{e_1+e_2+z_1+\alpha}{e_4}=\CC{e_1+e_2+z_1+\alpha}{e_5}=\CC{e_1+e_2+z_1+\alpha}{e_6}=+1$\\ 
\xrowht[()]{10pt}
&$\CC{e_1+e_2+z_1+\alpha}{b_1+\alpha}=\CC{e_1+e_2+z_1+\alpha}{\tilde{x}+\alpha}=\CC{e_1+e_2+z_1+\alpha}{z_2}=+1$\\
\hline
\xrowht[()]{10pt}
$e_1+e_2+e_3+\alpha $ &$\CC{e_1+e_2+e_3+\alpha}{e_4}=\CC{e_1+e_2+e_3+\alpha}{e_5}=\CC{e_1+e_2+e_3+\alpha}{e_6}=+1$\\ 
\xrowht[()]{10pt}
&$\CC{e_1+e_2+e_3+\alpha}{\tilde{x}+z_1+\alpha}=\CC{e_1+e_2+e_3+\alpha}{z_2}=+1$\\
\hline
\xrowht[()]{10pt}
$e_1+e_2+e_3+z_1+\alpha $ &$\CC{e_1+e_2+e_3+z_1+\alpha}{e_4}=\CC{e_1+e_2+e_3+z_1+\alpha}{e_5}=\CC{e_1+e_2+e_3+z_1+\alpha}{e_6}=+1$\\ 
\xrowht[()]{10pt}
&$\CC{e_1+e_2+e_3+z_1+\alpha}{\tilde{x}+\alpha}=\CC{e_1+e_2+e_3+z_1+\alpha}{z_2}=+1$\\
\hline
\end{tabular}
\caption{\label{StTachConds}\emph{Conditions on GGSO coefficients for survival of the on-shell PS tachyons for $\tilde{S}$--models}}
\end{table}

\FloatBarrier
\subsection{Massless Sectors} \label{sectors}
Now that we have a way to check that models are free of on--shell tachyons,
we can turn our attention to the massless sectors and their representations.
In Section 5 of \cite{so10tclass} a detailed analysis of the $SO(10)$ massless sectors was presented. In this section we will only mention the sectors from the $SO(10)$ level relevant for determining the observable phenomenology of a model. Otherwise we will focus on the new sectors to the PS models. 
\subsubsection{Observable Massless Sectors}

The spinorial $\mathbf{16}/\overline{\mathbf{16}}$ representations of $SO(10)$
arise from the 48 sectors (16 from each orbifold plane)
\begin{eqnarray}\label{spins1}
B_{pqrs}^{(1)} &=&  b_1 + pe_3 + qe_4 + re_5 + se_6
\nonumber \\ 
&=& \{\psi^{\mu},\chi^{1,2},(1-p)y^3\bar{y}^3,
pw^3\bar{w}^3,(1-q)y^4\bar{y}^4,qw^4\bar{w}^4,
 \nonumber\\
& & ~~~ (1-r)y^5\bar{y}^5,rw^5\bar{w}^5,(1-s)y^6\bar{y}^6,
sw^6\bar{w}^6,\bar{\eta}^{1},\bar{\psi}^{1,\ldots,5}\}
 \\
B_{pqrs}^{(2)} &=&   b_2 + pe_1 + qe_2 + re_5 + se_6\nonumber \\
B_{pqrs}^{(3)} &=&   b_3 + pe_1 + qe_2 + re_3 + se_4\nonumber
\end{eqnarray}
where $p,q,r,s = 0,1$ account for all combinations of shift vectors of
the internal fermions $\{y^i,w^i \ | \ \bar{y}^i,\bar{w}^i\}$. As in
previous classifications, we can now write down generic algebraic
equations to determine the number of $\mathbf{16}$ and $\overline{\mathbf{16}}$,
$N_{16}$ and $N_{\overline{16}}$, as a function of the GGSO coefficients.
To do this we first utilize the following projectors to determine which
of the 48 spinorial sectors survive
\begin{eqnarray}\label{SpinProjector}\nonumber
P^1_{pqrs} &=& \frac{1}{2^4}\prod_{i=1,2}\left(1-\CC{B^1_{pqrs}}{e_i}^* \right)\prod_{a=1,2}\left(1-\CC{B^1_{pqrs}}{z_a}^*\right)\\
P^2_{pqrs} &=& \frac{1}{2^4}\prod_{i=3,4}\left(1-\CC{
B^2_{pqrs}}{e_i}^*\right)\prod_{a=1,2}\left(1-\CC{B^2_{pqrs}}{z_a}^*\right)\\
\nonumber
P^3_{pqrs} &=& \frac{1}{2^4}\prod_{i=5,6}\left(1-\CC{B^3_{pqrs}}{e_i}^*\right)\prod_{a=1,2}\left(1-\CC{B^3_{pqrs}}{z_a}^*\right)
\end{eqnarray}
where we recall that the vector $z_2=\{\overline{\phi}^{5,6,7,8}\}$ is
the combination defined in eq. (\ref{z2}).
Then we define the chirality phases
\begin{eqnarray}
X^1_{pqrs} &=& -\CC{B^1_{pqrs}}{b_2+(1-r)e_5+(1-s)e_6}^* \nonumber\\
X^2_{pqrs} &=& -\CC{B^2_{pqrs}}{b_1+(1-r)e_5+(1-s)e_6}^*\\
X^3_{pqrs} &=& -\CC{B^3_{pqrs}}{b_1+(1-r)e_3+(1-s)e_4}^*\nonumber
\end{eqnarray}
to determine whether a sector will give rise to a $\mathbf{16}$ or a
$\overline{\mathbf{16}}$. With these definitions we can write compact
expressions for $N_{16}$ and $N_{\overline{16}}$
\begin{align}\label{16s}
\begin{split}
N_{16} &= \frac{1}{2}\sum_{\substack{A=1,2,3 \\ p,q,r,s=0,1}} 
P_{pqrs}^A\left(1 + X^A_{pqrs}\right) \\
N_{\overline{16}} &= \frac{1}{2}\sum_{\substack{A=1,2,3 \\ p,q,r,s=0,1}} 
P_{pqrs}^A\left(1 - X^A_{pqrs}\right). 
\end{split}
\end{align}
In order to describe the phenomenological properties of the models under consideration, we need to also consider what happens to the components of the $\mathbf{16}$/
$\overline{\mathbf{16}}$'s as the $SO(10)$ GUT is broken to the PS subgroup.

Recall that the basis vector $\alpha$ (\ref{alpha}) induces $SO(10)$ gauge symmetry breaking.
The spinorial $\mathbf{16}$/
$\overline{\mathbf{16}}$ representations of $SO(10)$ decompose under the 
$SO(6) \times SO(4)\equiv SU(4)\times SU(2)\times SU(2)$ gauge group as:
\begin{align}
\label{sp_deco}
\mathbf{16}=&(\mathbf{4}, \mathbf{2}, \mathbf{1})+(\mathbf{\bar{4}}, \mathbf{1}, \mathbf{2})=n_{4L}+n_{\bar{4}R}\\ 
\mathbf{\overline{16}}=&(\mathbf{\bar{4}}, \mathbf{2}, \mathbf{1})+(\mathbf{4}, \mathbf{1}, \mathbf{2})=n_{4R}+n_{\bar{4}L}
\nonumber
\end{align}
The GGSO projection under $\alpha$ determines which of the two possible PS 
representations a particular $\mathbf{16}$ or $\mathbf{\overline{16}}$ 
will fall into. Furthermore, we can make the connection to the Standard 
Model by noting that after Higgsing the $SU(2)_R$ gauge symmetry,
the $\mathbf{16}$ rep. decomposes as
\begin{align}
\label{sm_deco}
\mathbf{16} &= \ Q \left(\textbf{3} , \textbf{2} , \frac{1}{6}\right)  + L\left(1 , \textbf{2} , -\frac{1}{2}\right) \nonumber \\
&+ u^c\left(\overline{\textbf{3}} ,  \textbf{1} , -\frac{2}{3}\right) + d^c\left(\overline{\textbf{3}} ,  \textbf{1} , \frac{1}{3}\right) + e^c\left(\textbf{1} , \textbf{1},1\right) + \nu^c\left(\textbf{1} ,  \textbf{1} , 0\right)
\end{align}
under the SM gauge group.

A phenomenological issue arises due to the fact \ds--maps makes the would-be scalar degrees of
freedom $\tilde{S}+B^i_{pqrs}$ massive. In the supersymmetric models, the $\mathbf{16}$ and
$\mathbf{\overline{16}}$ include the scalar superpartners, in particular those of the $n_{\bar{4}R}$
and $n_{4R}$, which are used to break the intermediate PS gauge symmetry 
down to the Standard Model gauge group. Since there are no such states 
for our $\tilde{S}$--models and no suitable scalar exotic sectors or 
scalars from higher Kac--Moody level representations \cite{Reps}, 
the phenomenology of these models is seemingly incomplete. In the case of Standard--like
$\tilde{S}$--models such as \cite{stable}, however, the breaking of the 
additional $U(1)_{Z'}$ may be achieved by a scalar from the exotic sectors
\cite{exotics}. 
Despite this gap in the analysis of these \ds--models we continue with 
the analysis since our main aim here is to build up the tools for the 
classification for the \ds-- and $S$--models of Non-SUSY string models 
and to do a comparison of their key characteristics. 

The other key states for observable physics are obtained
from the vectorial $\mathbf{10}$ representations of $SO(10)$.
We note that the vector $\tilde{x}$ defined in (\ref{xtilde}) induces a map between the fermionic
spinorial sectors $B^{1,2,3}_{pqrs}$ and the bosonic vectorial sectors
\begin{eqnarray}\label{vects}
V_{pqrs}^{(1)} &=&  B_{pqrs}^{(1)}+\tilde{x}\nonumber \\
&=&b_2+b_3 + pe_3 + qe_4 + re_5 + se_6
\nonumber \\ 
&=& \{\chi^{3,4,5,6},(1-p)y^3\bar{y}^3,
pw^3\bar{w}^3,(1-q)y^4\bar{y}^4,qw^4\bar{w}^4,
 \nonumber\\
& & ~~~ (1-r)y^5\bar{y}^5,rw^5\bar{w}^5,(1-s)y^6\bar{y}^6,
sw^6\bar{w}^6,\bar{\eta}^{2,3}\}
 \\
V_{pqrs}^{(2)} &=&   B_{pqrs}^{(2)}+\tilde{x}\nonumber \\
V_{pqrs}^{(3)} &=&   B_{pqrs}^{(3)}+\tilde{x}\nonumber
\end{eqnarray}
The observable states will arise from these sectors when there is a right moving oscillator of
$\overline{\psi}^{a(*)}$, $a=1,...,5$. To determine the number of such observable vectorial sectors
we use the projectors
\begin{eqnarray}\label{VectProjector}
\nonumber R^{(1)}_{pqrs} &=& \frac{1}{2^4}\prod_{i=1,2}\left(1 + \CC{e_i}{V^{(1)}_{pqrs}}
\right)
\prod_{a=1,2}\left[1+ \CC{z_a}{V^{(1)}_{pqrs}}\right)\\
R^{(2)}_{pqrs} &=& \frac{1}{2^4}\prod_{i=3,4}\left(1+\CC{e_i}{V^{(2)}_{pqrs}}
\right)
\prod_{a=1,2}\left(1 + \CC{z_a}{V^{(2)}_{pqrs}}\right)\\
\end{eqnarray}
Then we can write the number of vectorial $\mathbf{10}$'s arising from these sectors as
\begin{eqnarray}
N_{10}=\sum_{\substack{A=1,2,3 \\ p,q,r,s=0,1 }}R^A_{pqrs}.
\end{eqnarray}
A GGSO projection with the vector $\alpha$ can then be used to induce doublet-triplet splitting in these models and tell us whether a particular $\mathbf{10}$ produces a bidoublet or a triplet.
In particular, the $\textbf{10}$ representation is decomposed under 
$SU(4) \times 
SU(2)_L \times SU(2)_R$ as:
\begin{align}
\textbf{10} = D(\mathbf{6},\mathbf{1},\mathbf{1})+h(\mathbf{1},\mathbf{2},\mathbf{2})
\label{vec_deco}
\end{align}
where the colored triplets are generated by the 
$\overline{\psi}^{1,2,3}_{1/2}/\overline{\psi}^{* 1,2,3}_{1/2}$ and the bi--doublet is generated by 
$\overline{\psi}^{4,5}_{1/2}/\overline{\psi}^{*4,5}_{1/2}$ oscillators. We can write the number of bidoublets, $n_h$, and number of triplets, $n_6$, algebraically as
\begin{align}\label{DoubTrip}
\begin{split}
n_{h} &= \frac{1}{2}\sum_{\substack{A=1,2,3 \\ p,q,r,s=0,1}} 
R_{pqrs}^A\left(1 - \CC{V^A_{pqrs}}{\alpha} \right) \\
n_6 &= \frac{1}{2}\sum_{\substack{A=1,2,3 \\ p,q,r,s=0,1}} 
R_{pqrs}^A\left(1 + \CC{V^A_{pqrs}}{\alpha}\right). 
\end{split}
\end{align}
\subsubsection{Pati--Salam Exotics}
Having reviewed the key elements of the observable massless sectors, we can turn our attention to the
new features of the massless spectrum due to the PS vector $\alpha$. Such sectors are
exotic in the sense they transform under both the hidden and observable gauge groups and carry
fractional electric charge.

For the purposes of our analysis we will only classify the fermionic exotic sectors and ensure that
any potentially viable models are checked for the absence of chiral exotics. In a classification of
supersymmetric models it would be sufficient to classify only the fermionic exotic sectors and then
know that the equivalent scalar exotic sectors will be in a model with $+S$. However, in our
non-supersymmetric models we have no control over how many scalar exotics will arise in a model when
we check the fermionic exotic sectors. We note that the scalar exotics can always gain heavy mass 
since they are not chiral. 

First of all we note two fermionic vectorial exotic sectors
\beq \label{ExotVecs}
\tilde{V}_{\text{PS}}= 
\begin{Bmatrix}
\{\bar{\lambda}^i\} \ket{\tilde{S}+z_1+\alpha}, \ \ \ \ 
\{\bar{\lambda}^i\} \ket{\tilde{S}+z_1+z_2+\alpha}
\end{Bmatrix}.
\eeq 
Then we have in total 196 fermionic spinorial exotics. Of these, 96 arise from sectors transforming as  $(\mathbf{4},\mathbf{2},\mathbf{1}), (\mathbf{4},\mathbf{1},\mathbf{2}), (\overline{\mathbf{4}},\mathbf{2},\mathbf{1})$ or $(\overline{\mathbf{4}},\mathbf{1},\mathbf{2})$ where $\mathbf{4}$  and $\overline{\mathbf{4}}$ are spinorial and anti--spinorial representations of the observable $SU(4)$, respectively, and the $\mathbf{2}$ are doublet representations of either the first or second hidden $SO(4)$ factor. Explicitly, these sectors are
\begin{eqnarray}\label{fermExot1}
E_{pqrs}^{(1)} &=&  B^{(1)}_{pqrs}+\alpha
\nonumber \\ 
&=& \{\psi^{\mu},\chi^{1,2},(1-p)y^3\bar{y}^3,
pw^3\bar{w}^3,(1-q)y^4\bar{y}^4,qw^4\bar{w}^4,
 \nonumber\\
& & ~~~ (1-r)y^5\bar{y}^5,rw^5\bar{w}^5,(1-s)y^6\bar{y}^6,
sw^6\bar{w}^6,\bar{\eta}^{1},\bar{\psi}^{1,2,3},\bar{\phi}^{1,2}\} \nonumber\\
E_{pqrs}^{(2)} &=&   B^{(2)}_{pqrs}+\alpha \nonumber\\
E_{pqrs}^{(3)} &=&   B^{(3)}_{pqrs}+\alpha\\
E_{pqrs}^{(4)} &=&  B^{(1)}_{pqrs}+z_1+\alpha
\nonumber \\ 
&=& \{\psi^{\mu},\chi^{1,2},(1-p)y^3\bar{y}^3,
pw^3\bar{w}^3,(1-q)y^4\bar{y}^4,qw^4\bar{w}^4,
 \nonumber\\
& & ~~~ (1-r)y^5\bar{y}^5,rw^5\bar{w}^5,(1-s)y^6\bar{y}^6,
sw^6\bar{w}^6,\bar{\eta}^{1},\bar{\psi}^{1,2,3},\bar{\phi}^{3,4}\} \nonumber\\
E_{pqrs}^{(5)} &=&   B^{(2)}_{pqrs}+z_1+\alpha \nonumber\\
E_{pqrs}^{(6)} &=&   B^{(3)}_{pqrs}+z_1+\alpha\nonumber
\end{eqnarray}
A further 96 exotic sectors arise from
\begin{eqnarray}\label{fermExot2}
E_{pqrs}^{(7)} &=&  \tilde{S}+V^{(1)}_{pqrs}+z_1+\alpha
\nonumber \\ 
&=& \{\psi^{\mu},\chi^{1,2},(1-p)y^3\bar{y}^3,
pw^3\bar{w}^3,(1-q)y^4\bar{y}^4,qw^4\bar{w}^4,
 \nonumber\\
& & ~~~ (1-r)y^5\bar{y}^5,rw^5\bar{w}^5,(1-s)y^6\bar{y}^6,
sw^6\bar{w}^6,\bar{\eta}^{2,3},\bar{\psi}^{4,5},\bar{\phi}^{5,6}\}\nonumber\\
E_{pqrs}^{(8)} &=&  \tilde{S}+ V^{(2)}_{pqrs}+z_1+\alpha \nonumber\\
E_{pqrs}^{(9)} &=&  \tilde{S}+ V^{(3)}_{pqrs}+z_1+\alpha\\
E_{pqrs}^{(10)} &=& \tilde{S}+ V^{(1)}_{pqrs}+z_1+z_2+\alpha
\nonumber \\ 
&=& \{\psi^{\mu},\chi^{1,2},(1-p)y^3\bar{y}^3,
pw^3\bar{w}^3,(1-q)y^4\bar{y}^4,qw^4\bar{w}^4,
 \nonumber\\
& & ~~~ (1-r)y^5\bar{y}^5,rw^5\bar{w}^5,(1-s)y^6\bar{y}^6,
sw^6\bar{w}^6,\bar{\eta}^{2,3},\bar{\psi}^{4,5},\bar{\phi}^{7,8}\}\nonumber\\
E_{pqrs}^{(11)} &=&  \tilde{S}+ V^{(2)}_{pqrs}+z_1+z_2+\alpha \nonumber\\
E_{pqrs}^{(12)} &=& \tilde{S}+  V^{(3)}_{pqrs}+z_1+z_2+\alpha\nonumber
\end{eqnarray}
which give representations  $((\mathbf{2},\mathbf{1}),(\mathbf{2},\mathbf{1})), ((\mathbf{2},\mathbf{1}),(\mathbf{1},\mathbf{2})),((\mathbf{1},\mathbf{2}),(\mathbf{1},\mathbf{2}))$ or $((\mathbf{1},\mathbf{2}),(\mathbf{2},\mathbf{1}))$ under $SU(2)_L\times SU(2)_R\times SO(4)_{3/4}$. 

Finally there are 4 additional fermionic spinorial exotics from the sectors: 
\beq \label{rogueExots}
\begin{Bmatrix}
\tilde{S}+\alpha, \ \ \ 
\tilde{S}+z_2+\alpha, \ \ \  
\tilde{x}+\alpha, \ \ \ \tilde{x}+z_1+\alpha
\end{Bmatrix}
\eeq 
which transform as a doublet under one of the observable $SU(2)$ factors and doublets under three of the hidden $SO(4)$ gauge factors. 

In regard to phenomenology, we are interested in ensuring that there are no chiral exotics in our models, which is why we look at the fermionic exotic sectors only. In order to check this we can classify the exotics in eq. (\ref{fermExot1}) with the numbers $n_4,n_{\bar{4}}$  counting the number of $\mathbf{4}/\mathbf{\bar{4}}$'s under the $SU(4)$ gauge factor and count the numbers $n_{2L},n_{2R}$ of $(\mathbf{1},\mathbf{2})/(\mathbf{2},\mathbf{1})$'s from equations (\ref{fermExot2}) and (\ref{rogueExots}) under the $SU(2)\times SU(2)$ gauge factors.

\section{Fertility Methodology in $\tilde{S}$--Models}\label{FertMeth}
The application of the fertility methodology was employed very successfully in the case of the
supersymmetric standard--like models in \cite{slmclass} and left-right symmetric models in
\cite{ferlrs}. The methodology was particularly useful in these cases due to the scarcity of
phenomenologically viable vacua, which was largely due to the abundance of exotic sectors in these
models. The key notable feature of our $\tilde{S}$--models which make them ripe for a fertility
methodology is the abundance of tachyonic models. It was noted in \cite{so10tclass} that the
probability of tachyon--free $SO(10)$ models was $\sim 5\times 10^{-3}$, and we can similarly find
that the probability for PS $\tilde{S}$--models to be tachyon--free is $\sim 2 \times 10^{-3}$,
\textit{i.e. } we have to filter out all but around 1 in 500 GGSO configurations.

The fertility methodology involves splitting the full parameter space of, in this case, 
PS models into two components $\Pi=\Pi_{SO(10)}\times \Pi_{\alpha}$ and performing a 
classification in two steps. The first step is to take the $SO(10)$ subspace $\Pi_{SO(10)}$, 
which is the space of $2^{66}$ independent GGSO phase configurations for the 12 basis 
vectors (\ref{basisSt}), and checking for a set of phenomenological conditions 
which can solely be evaluated within this subspace. These conditions we 
call `fertility conditions' and the $SO(10)$ models satisfying them `fertile cores'. 
The fertility conditions we impose in our analysis of the \ds--models are listed below. 

Once we have these $SO(10)$ fertile cores, we perform the second step which is to 
evaluate them in the space $\Pi_{\alpha}$, which means iterating over all independent 
choices of the GGSO phases involving $\alpha$ around these cores. We then collect statistics 
for these PS models and find a much increased efficiency in finding models satisfying 
phenomenological constraints owing to this fertility methodology, compared with a random 
classification in the full space of PS models. 

\subsection{$\tilde{S}$ Fertility Conditions} \label{StFertConds}
The set of fertility conditions we will use to derive 
fertile $SO(10)$ $\tilde{S}$ cores can be listed as follows: 
\begin{enumerate}
\item[1)] Absence of the tachyonic sectors: 
\beq \label{z1TachCondSt}
z_1, \ e_i+z_1, \ e_i+e_j+z_1,\ e_i+e_j+e_k+z_1, \ \ \ \ i\neq j \neq k =1,2,3,4,5,6
\eeq 
using the set of projection conditions delineated in Section 4 of \cite{so10tclass}. 
\item[2)] Constraints on $SO(10)$ spinorial states
\begin{align}\label{fertility1}
 n_{4L}-n_{\bar{4}L}  =  n_{\bar{4}R} -n_{4R}\geq 6 \ ,\ n_{\bar{4}R}^F > 6 .
\end{align}
The first condition results in a high likelihood of having (at least) 3 fermion families and the second condition is used as a fermionic analogy to the condition for a Heavy Higgs, which would be $n_{\bar{4}R}^B > 1$ if the \ds--models had the scalar partners of $B^{(1,2,3)}$. 
\item[3)] For the presence of a SM higgs, \textit{i.e.} $n_h\geq 1$, and a $D(6,1,1)$, \textit{i.e.} $n_6 \geq 1$, we can impose
\begin{align}\label{fertility2}
N_{10}\geq 2
\end{align}
at the $SO(10)$ level. 
\item[4)] Presence of a top quark mass coupling (TQMC) which amounts to fixing the following GGSO coefficients 
\begin{eqnarray}\label{tqmcSO10}
& &\CC{b_1}{e_1} = \CC{b_1}{e_2} = 
\CC{b_1}{z_1} = \CC{b_1}{z_2} = -1\ ,\nonumber\\
& &\CC{b_2}{e_3} = \CC{b_2}{e_4} = 
\CC{b_2}{z_1} = \CC{b_2}{z_2} = -1\ ,\\
& & \CC{e_5}{b_1}\CC{e_5}{b_2}=\CC{e_6}{b_1}\CC{e_6}{b_2}=1\nonumber\\
& &\CC{b_1}{b_2} =  -\CC{e_5}{b_1} \CC{e_6}{b_1}\nonumber,
\end{eqnarray}
at the $SO(10)$ level.

Since $z_2$ is a combination of basis vectors we can rewrite the conditions $\CC{b_1}{z_2} =\CC{b_2}{z_2} = -1$ in terms of the GGSO phases between basis vectors using the equations in (\ref{tqmcSO10}) and ABK rules, choosing to fix the phases
\begin{align}
\begin{split}\label{z2TQMfixes}
  \CC{b_1}{b_3}&=-\CC{b_1}{e_3}\CC{b_1}{e_4}\\
  \CC{b_2}{b_3}&=-\CC{b_2}{e_1}\CC{b_2}{e_2}.
 \end{split}
\end{align}
\end{enumerate}
Having listed the conditions 1)-4), we will now explain them in more detail. 

In regard to condition 1), we should note that these 42 tachyonic sectors (\ref{z1TachCondSt})
involving $z_1$ are the only $SO(10)$ tachyonic sectors we can use as a fertility condition, since
the others can be projected by an $\alpha$ GGSO projection as mentioned in Section \ref{tsa}. For
example, we could have an $SO(10)$ core with a spinorial tachyon from the $z_2$ sector which is in
fact absent at the PS level due to $\CC{z_2}{\alpha}=-1$. However, the $z_1$ tachyonic sectors must
be projected at the $SO(10)$ level as a necessary condition on the absence of all tachyons at the PS
level. 

In regard to the second part of condition 2), 
having a $n_{4R}^B-n_{\bar{4}R}^B$ present we typically call having a heavy Higgs, where we use the
$B$ superscript to make it clear that this is a bosonic (scalar) sector.
However, as already mentioned, the \ds--models have no such scalar components and so we cannot
implement this condition. Instead we implement a condition on having the fermionic representations
$n_{\bar{4}R},n_{4R}$ in the spectrum in order to draw a parallel with the Heavy Higgs condition
which we can implement for the $S$--models analysed in the next section.
We further must note that the condition $n_6\geq 1$ in 3), relates to the so--called Missing Partner
Mechanism of Pati--Salam models \cite{antonandleon} 
for which we also require the $n_{4R}^{(B)},n_{\bar{4}R}^{(B)}$.
Again, since we have no Heavy Higgs for the \ds--models we implement this $n_6\geq 1$ condition purely
to draw the parallel to the methodology we can apply for the $S$--models. In analogy with the Missing
Partner Mechanism we will say that the requirement of $n_6\geq 1$ is required for a `Heavy Triplet
Constraint'.


The conditions on the GGSO phases from 4) in equations (\ref{tqmcSO10}) and (\ref{z2TQMfixes}) for the existence of the TQMC is a necessary condition for the presence of a TQMC at the PS level and can be made into a sufficient condition by fixing the phases
\begin{align} \label{tqmcPS}
\CC{b_1}{\alpha} = -\CC{b_2}{\alpha} = -1
\end{align}
at the PS level. The derivation of these results in a supersymmetric PS construction can be found in ref. \cite{tqmc}. It is derived through choosing $B^1_{0000}=b_1$ and $B^2_{0000}=b_2$ to survive the GGSO projections (\ref{SpinProjector}), along with a bidoublet Higgs from the sector $V^3_{0000}=b_1+b_2$ surviving the projector (\ref{DoubTrip}). Furthermore, using (\ref{16s}) we can ensure $b_1$ and $b_2$ give rise to $\mathbf{16}$'s such that $X^1_{0000}=X^2_{0000}=1$ through the condition (\ref{tqmcSO10}). Then the GGSO phase conditions (\ref{tqmcPS}) ensures that $b_1$ generates a $n_{4L} \ (\mathbf{4},\mathbf{2},\mathbf{1})$, $b_2$ generates a $n_{\bar{4}R} \ (\mathbf{\bar{4}},\mathbf{1},\mathbf{2})$ and then $b_1+b_2$ generates the $h \ (\mathbf{1},\mathbf{2},\mathbf{2})$. We thus obtain the TQMC via the term: 
\begin{align} \label{tqmcSPot}
n_{4L}^{(1) F} n_{\bar{4}R}^{(2) F} h^{(3)B}=Qu^c H^u+Q d^c H^d+L e^c H^d+L \nu^c H^u.
\end{align}

Since we have guaranteed that a bidoublet Higgs arises from the sector $b_1+b_2$ we can note this overlaps with condition 3) since we automatically have $N_{10}\geq 1$. 

Having defined the fertility methodology and the details of the construction of PS
$\tilde{S}$--models, in the next section we present the construction of the non-supersymmetric
$S$--models which descend from the non--tachyonic $SO(16)\times SO(16)$ 10D string. The results of the
classification of PS $\tilde{S}$--models and $S$--models are given in Section \ref{results}.

\section{$SO(16)\times SO(16)$--Derived 4D Models}
\label{PSNormS}
We now turn to look at what we will refer to as the $S$--models, which are the non-supersymmetric
class of models descending from the $SO(16)\times SO(16)$ heterotic string. We can employ the same
basis as used for the classification of supersymmetric PS models in \cite{acfkr}
\begin{align}\label{basisS}
\mathds{1}&=\{\psi^\mu,\
\chi^{1,\dots,6},y^{1,\dots,6}, w^{1,\dots,6}\ | \ \overline{y}^{1,\dots,6},\overline{w}^{1,\dots,6},
\overline{\psi}^{1,\dots,5},\overline{\eta}^{1,2,3},\overline{\phi}^{1,\dots,8}\},\nonumber\\
S&=\{{\psi^\mu},\chi^{1,\dots,6} \},\nonumber\\
{e_i}&=\{y^{i},w^{i}\; | \; \overline{y}^{i},\overline{w}^{i}\}, 
\ \ \ i=1,...,6
\nonumber\\
{b_1}&=\{\chi^{34},\chi^{56},y^{34},y^{56}\; | \; \overline{y}^{34},
\overline{y}^{56},\overline{\eta}^1,\overline{\psi}^{1,\dots,5}\},\\
{b_2}&=\{\chi^{12},\chi^{56},y^{12},y^{56}\; | \; \overline{y}^{12},
\overline{y}^{56},\overline{\eta}^2,\overline{\psi}^{1,\dots,5}\},\nonumber\\
z_1&=\{\overline{\phi}^{1,\dots,4}\},\nonumber\\
z_2&=\{\overline{\phi}^{5,\dots,8}\},\nonumber\\
\alpha&=\{\overline{\psi}^{4,5},\overline{\phi}^{1,2} \}
\nonumber
\end{align}
We will also make regular use of the combination
\beq 
x=1+S+\sum_{i=1}^6 e_i +\sum_{k=1}^2 z_k=\{\bar{\psi}^{1,...,5},\bar{\eta}^{1,2,3}\}.
\eeq 
The non--SUSY models are those in which the gravitino is projected from the $S$ 
massless sector. This means, in other words, that SUSY is broken by a GGSO phase in these $S$--models.
Therefore, the total space of non--supersymmetric models is the total space 
of all PS models which is $2^{13(13-1)/2}=2^{78}$ models\footnote{The phase
$\CC{\mathds{1}}{\mathds{1}}$ can, without loss of generality, be fixed to $+1$ or $-1$, which just
leaves the upper triangle of the GGSO phase matrix as free phases, with the rest fixed by modular
invariance.} minus the space of supersymmetric configurations. The necessary and sufficient condition
for the presence of supersymmetry is the fixing of the 9 GGSO phases
\beq \label{SUSYphases}
\CC{S}{e_i}=\CC{S}{z_n}=\CC{S}{\alpha}=-1,
\eeq 
with $i=1,...,6$, $m=1,2$ and  $n=1,2$. These phase choices ensure the gravitino is not projected. Therefore the total space of non--SUSY PS models has size 
$2^{78}-2^{78-9}\sim 3 \times 10^{23}$. 

The untwisted sector gauge vector
bosons for these models generate the gauge group
\beq 
SO(6)\times SO(4) \times  U(1)_1\times U(1)_2\times U(1)_3\times SO(4)^2\times SO(8).
\eeq 
which only differ from the $\tilde{S}$--models in that the gauge bosons generating the $SO(8)$ hidden group factor $\psi^{\mu}\bar{\phi}^c\bar{\phi}^d\ket{NS}$, $c,d=5,6,7,8$, are unbroken unlike in the case of the $\tilde{S}$ projection. Additional gauge bosons arise from enhanced sectors which are exactly those listed in the supersymmetric case from Section 3.1 of \cite{acfkr}. 
\subsection{Analysis of Tachyonic Sectors}\label{abcd}
The tachyonic sectors for the $S$--models are identical to the $\tilde{S}$ case, however the projection/survival conditions differ. Having $S$ in our basis ($\ref{basisS}$), means it can be used as a possible projector of all the tachyonic sectors depending on the GGSO phases involving $S$. Of course, in the SUSY case where eq. (\ref{SUSYphases}) holds, all the tachyonic sectors are automatically projected by the $S$ projection, which is the origin of SUSY models being tachyon--free in the free fermionic construction. 

Due to this special property of $S$, in terms of computational efficiency it makes sense to implement the $S$ projection first on all the tachyonic sectors and then only apply the further GGSO projections to sectors surviving $S$. Concretely, for a generic tachyonic sector $\xi$, in which eq. (\ref{onshelltach}) holds, we can first check 
\beq 
\CC{\xi}{S}=\begin{cases}
+1 \ \ \  \ \text{survives}\\
-1 \ \ \ \ \text{projected}
\end{cases}
\eeq 
and if $\xi$ survives this then we can move on to check the other GGSO projections for its survival/absence. 

The conditions for the survival of all $SO(10)$ tachyonic sectors are detailed in Tables \ref{NormSSO10} and \ref{NormSSO10Vect} before the PS tachyonic sectors survival conditions are explicated in Table \ref{NormSal}. 
\begin{table}[!ht] 
\centering
\small
\setlength{\tabcolsep}{1.5pt}
\begin{tabular}{|c|l|}
\hline 
\xrowht[()]{8pt}
\textbf{Tachyonic }&\textbf{Survival conditions}\\
\xrowht[()]{10pt}
\textbf{Sector}&\\
\hline \xrowht[()]{10pt}
$z_1$&$\CC{z_1}{e_1}=\CC{z_1}{e_2}=\CC{z_1}{e_3}=\CC{z_1}{e_4}=\CC{z_1}{e_5}=\CC{z_1}{e_6}=+1$\\
\xrowht[()]{10pt}
&$\CC{z_1}{S}=\CC{z_1}{z_2}=\CC{z_1}{b_1}=\CC{z_1}{b_2}=\CC{z_1}{x}=+1$ \\ [0.75ex] 
\hline

\xrowht[()]{10pt}
$e_1+z_1 $ &$\CC{e_1+z_1}{e_2}=\CC{e_1+z_1}{e_3}=\CC{e_1+z_1}{e_4}=\CC{e_1+z_1}{e_5}=\CC{e_1+z_1}{e_6}=+1$\\ 
\xrowht[()]{10pt}
&$\CC{e_1+z_1}{S}=\CC{e_1+z_1}{b_1}=\CC{e_1+z_1}{z_2}=\CC{e_1+z_1}{x}=+1$\\
\hline

\xrowht[()]{10pt}
$e_1+e_2+z_1 $ &$\CC{e_1+e_2+z_1}{e_3}=\CC{e_1+e_2+z_1}{e_4}=\CC{e_1+e_2+z_1}{e_5}=\CC{e_1+e_2+z_1}{e_6}=+1$\\ 
\xrowht[()]{10pt}
&$\CC{e_1+e_2+z_1}{S}=\CC{e_1+e_2+z_1}{b_1}=\CC{e_1+e_2+z_1}{x}=\CC{e_1+e_2+z_1}{z_2}=+1$\\
\hline
\xrowht[()]{10pt}
$e_1+e_2+e_3+z_1 $ &$\CC{e_1+e_2+e_3+z_1}{e_4}=\CC{e_1+e_2+e_3+z_1}{e_5}=\CC{e_1+e_2+e_3+z_1}{e_6}=+1$\\ 
\xrowht[()]{10pt}
&$\CC{e_1+e_2+e_3+z_1}{S}=\CC{e_1+e_2+e_3+z_1}{x}=\CC{e_1+e_2+e_3+z_1}{z_2}=+1$\\
\hline
\xrowht[()]{10pt}
$z_2$&$\CC{z_2}{e_1}=\CC{z_2}{e_2}=\CC{z_2}{e_3}=\CC{z_2}{e_4}=\CC{z_2}{e_5}=\CC{z_2}{e_6}=+1$\\ 
\xrowht[()]{10pt}
&$\CC{z_2}{S}=\CC{z_2}{\alpha}=\CC{z_2}{z_1}=\CC{z_2}{b_1}=\CC{z_2}{b_2}=\CC{z_2}{x}=+1$\\
\hline
\xrowht[()]{10pt}
$e_1+z_2 $ &$\CC{e_1+z_2}{e_2}=\CC{e_1+z_2}{e_3}=\CC{e_1+z_2}{e_4}=\CC{e_1+z_2}{e_5}=\CC{e_1+z_2}{e_6}=+1$\\ 
\xrowht[()]{10pt}
&$\CC{e_1+z_2}{S}=\CC{e_1+z_2}{\alpha}=\CC{e_1+z_2}{b_1}=\CC{e_1+z_2}{x}=\CC{e_1+z_2}{z_1}=+1$\\
\hline
\xrowht[()]{10pt}
$e_1+e_2+z_2 $ &$\CC{e_1+e_2+z_2}{e_3}=\CC{e_1+e_2+z_2}{e_4}=\CC{e_1+e_2+z_2}{e_5}=\CC{e_1+e_2+z_2}{e_6}=+1$\\ 
\xrowht[()]{10pt}
&$\CC{e_1+e_2+z_2}{S}=\CC{e_1+e_2+z_2}{\alpha}=\CC{e_1+e_2+z_2}{b_1}=\CC{e_1+e_2+z_2}{x}=\CC{e_1+e_2+z_2}{z_1}=+1$\\
\hline
\xrowht[()]{10pt}
$e_1+e_2+e_3+z_2 $ &$\CC{e_1+e_2+e_3+z_2}{e_4}=\CC{e_1+e_2+e_3+z_2}{e_5}=\CC{e_1+e_2+e_3+z_2}{e_6}=+1$\\ 
\xrowht[()]{10pt}
&$\CC{e_1+e_2+e_3+z_2}{S}=\CC{e_1+e_2+e_3+z_2}{\alpha}=\CC{e_1+e_2+e_3+z_2}{x}=\CC{e_1+e_2+e_3+z_2}{z_1}=+1$\\
\hline
\end{tabular}
\caption{\label{NormSSO10}\emph{Conditions on GGSO coefficients for survival of spinorial $SO(10)$ level on-shell tachyons for $S$--models. Only $e_1,e_1+e_2,e_1+e_2+e_3$ combinations are detailed but other combinations of $e_i$'s are similar. }}
\end{table}
\begin{table}[!ht] 
\centering
\small
\setlength{\tabcolsep}{1.5pt}
\begin{tabular}{|c|l|}
\hline \xrowht[()]{8pt}
\textbf{Tachyonic }&\textbf{Survival conditions}\\
\xrowht[()]{10pt}
\textbf{Sector}&\\
\hline 
\xrowht[()]{10pt}
$\{ \bar{\lambda}\}\ket{e_1} $ & $ S_O=\Big\{\CC{e_1}{e_2},\CC{e_1}{e_3},\CC{e_1}{e_4},\CC{e_1}{e_5},\CC{e_1}{e_6},\CC{e_1}{z_1},\CC{e_1}{z_2},\CC{e_1}{\alpha},\CC{e_1}{b_1},\CC{e_1}{x}\Big\}$\\
\xrowht[()]{10pt}
$\bar{\lambda}=\bar{\psi}^{1,2,3}/\bar{\eta}^{1}$&$ \ \CC{e_1}{S}=1 \ \text{ and } \ \#(x \in S_O | x=-1)=2 \ \text{s.t. } \CC{e_1}{b_1}=\CC{e_1}{x}=-1$ \  \\ 
\xrowht[()]{10pt}
$\bar{\lambda}=\bar{\phi}^{1,2}$&$\text{ \textbf{or}  } \ \CC{e_1}{S}=1 \ \text{ and } \ \#(x \in S_O | x=-1)=2 \ \text{s.t. } \CC{e_1}{z_1}=\CC{e_1}{\alpha}=-1$ \  \\ 
\xrowht[()]{10pt}
$\bar{\lambda}=\bar{y}^{3,4,5,6}$&$\text{ \textbf{or } } \ \CC{e_1}{S}=1 \ \text{ and } \ \#(x \in S_O | x=-1)=2 \ \text{s.t. } \CC{e_1}{b_1}=\CC{e_1}{e_{3,4,5,6}}=-1$ \  \\
\xrowht[()]{10pt}
$\bar{\lambda}=\bar{\psi}^{4,5}$&$\text{ \textbf{or } } \ \CC{e_1}{S}=1 \ \text{ and } \ \#(x \in S_O | x=-1)=3 \ \text{s.t. } \CC{e_1}{b_1}=\CC{e_1}{x}=\CC{e_1}{\alpha}=-1$ \  \\
$\bar{\lambda}=\text{ else}$ &$\text{ \textbf{or } }\CC{e_1}{S}=1 \ \text{ and } \ \#(x \in S_O | x=-1)=1$ \  \\ 
[0.75ex]
\hline
\xrowht[()]{10pt}
$\{\bar{\lambda}\} \ket{e_{12}}  $ &$S_O=\Big\{\CC{e_{12}}{e_3},\CC{e_{12}}{e_4},\CC{e_{12}}{e_5},\CC{e_{12}}{e_6},\CC{e_{12}}{z_1},\CC{e_{12}}{z_2},\CC{e_{12}}{\alpha},\CC{e_{12}}{b_1},\CC{e_{12}}{x}\Big\}$\\
\xrowht[()]{10pt}
$\bar{\lambda}=\bar{\psi}^{1,2,3}/\bar{\eta}^{1}$&$\ \CC{e_{12}}{S}=1 \ \text{ and } \ \#(x \in S_O | x=-1)=2 \ \text{s.t. } \CC{e_{12}}{b_1}=\CC{e_{12}}{x}=-1$ \  \\ 
\xrowht[()]{10pt}
$\bar{\lambda}=\bar{\phi}^{1,2}$&$\text{ \textbf{or}  } \ \CC{e_{12}}{S}=1 \ \text{ and } \ \#(x \in S_O | x=-1)=2 \ \text{s.t. } \CC{e_{12}}{z_1}=\CC{e_{12}}{\alpha}=-1$ \  \\ 
\xrowht[()]{10pt}
$\bar{\lambda}=\bar{y}^{3,4,5,6}$&$\text{ \textbf{or } } \ \CC{e_{12}}{S}=1 \ \text{ and } \ \#(x \in S_O | x=-1)=2 \ \text{s.t. } \CC{e_{12}}{b_1}=\CC{e_{12}}{e_{3,4,5,6}}=-1$ \  \\
\xrowht[()]{10pt}
$\bar{\lambda}=\bar{\psi}^{4,5}$&$\text{ \textbf{or } } \ \CC{e_{12}}{S}=1 \ \text{ and } \ \#(x \in S_O | x=-1)=3 \ \text{s.t. } \CC{e_{12}}{b_1}=\CC{e_{12}}{x}=\CC{e_{12}}{\alpha}=-1$ \  \\
\xrowht[()]{10pt}
$\bar{\lambda}=\text{ else}$&$\text{ \textbf{or } }\CC{e_{12}}{S}=1 \ \text{ and } \ \#(x \in S_O | x=-1)=1$ \  \\ [0.75ex]
\hline
\xrowht[()]{10pt}
$\{\bar{\lambda}\}\ket{e_{123}}  $ &$S_O=\Big\{\CC{e_{123}}{e_4},\CC{e_{123}}{e_5},\CC{e_{123}}{e_6},\CC{e_{123}}{z_1},\CC{e_{123}}{z_2},\CC{e_{123}}{\alpha},\CC{e_{123}}{x}\Big\}$\\
\xrowht[()]{10pt}
$\bar{\lambda}=\bar{\psi}^{4,5}$&$ \ \CC{e_{123}}{S}=1 \ \text{ and } \ \#(x \in S_O | x=-1)=2 \ \text{s.t. } \CC{e_{123}}{\alpha}=\CC{e_{123}}{x}=-1$ \  \\ 
\xrowht[()]{10pt}
$\bar{\lambda}=\bar{\phi}^{1,2}$&$\text{ \textbf{or}  } \ \CC{e_{123}}{S}=1 \ \text{ and } \ \#(x \in S_O | x=-1)=2 \ \text{s.t. } \CC{e_{123}}{z_1}=\CC{e_{123}}{\alpha}=-1$ \  \\ 
\xrowht[()]{10pt}
$\bar{\lambda}=\text{ else}$&$\text{ \textbf{or}  }\CC{e_{123}}{S}=1 \ \text{ and } \ \#(x \in S_O | x=-1)=1$ \  \\ 
[0.75ex]
\hline
\end{tabular}
\caption{\label{NormSSO10Vect}\emph{Conditions on GGSO coefficients for the survival of vectorial $SO(10)$ level on-shell tachyons in $S$--models. Only $e_1,e_{12}:=e_1+e_2$ and $e_{123}:=e_1+e_2+e_3$ combinations are detailed but other combinations of $e_i$'s are similar.}}
\end{table}

\begin{table}[!ht] 
\centering
\small
\setlength{\tabcolsep}{1.5pt}
\begin{tabular}{|c|l|}
\hline 
\textbf{Tachyonic sector}&\textbf{Survival conditions}\\
\hline \xrowht[()]{10pt}
$\alpha$&$\CC{\alpha}{e_1}=\CC{\alpha}{e_2}=\CC{\alpha}{e_3}=\CC{\alpha}{e_4}=\CC{\alpha}{e_5}=\CC{\alpha}{e_6}=+1$\\
\xrowht[()]{10pt}
&$\CC{\alpha}{S}=\CC{\alpha}{z_2}=\CC{\alpha}{b_1+z_1+\alpha}=\CC{\alpha}{b_2+z_1+\alpha}=+1$ \\ [0.75ex] 
\hline
\xrowht[()]{10pt}
$z_1+\alpha$&$\CC{z_1+\alpha}{e_1}=\CC{z_1+\alpha}{e_2}=\CC{z_1+\alpha}{e_3}=\CC{z_1+\alpha}{e_4}=\CC{z_1+\alpha}{e_5}=\CC{z_1+\alpha}{e_6}=+1$\\ 
\xrowht[()]{10pt}
&$\CC{z_1+\alpha}{S}=\CC{z_1+\alpha}{z_2}=\CC{z_1+\alpha}{b_1+\alpha}=\CC{z_1+\alpha}{b_2+\alpha}=+1$\\
\hline
\xrowht[()]{10pt}
$e_1+\alpha $ &$\CC{e_1+\alpha}{e_2}=\CC{e_1+\alpha}{e_3}=\CC{e_1+\alpha}{e_4}=\CC{e_1+\alpha}{e_5}=\CC{e_1+\alpha}{e_6}=+1$\\ 
\xrowht[()]{10pt}
&$\CC{e_1+\alpha}{S}=\CC{e_1+\alpha}{b_1+z_1+\alpha}=\CC{e_1+\alpha}{\tilde{x}+z_1+\alpha}=\CC{e_1+\alpha}{z_2}=+1$\\
\hline
\xrowht[()]{10pt}
$e_1+z_1+\alpha $ &$\CC{e_1+z_1+\alpha}{e_2}=\CC{e_1+z_1+\alpha}{e_3}=\CC{e_1+z_1+\alpha}{e_4}=\CC{e_1+z_1+\alpha}{e_5}=\CC{e_1+z_1+\alpha}{e_6}=+1$\\ 
\xrowht[()]{10pt}
&$\CC{e_1+z_1+\alpha}{S}=\CC{e_1+z_1+\alpha}{b_1+\alpha}=\CC{e_1+z_1+\alpha}{\tilde{x}+\alpha}=\CC{e_1+z_1+\alpha}{z_2}=+1$\\
\hline
\xrowht[()]{10pt}
$e_1+e_2+\alpha $ &$\CC{e_1+e_2+\alpha}{e_3}=\CC{e_1+e_2+\alpha}{e_4}=\CC{e_1+e_2+\alpha}{e_5}=\CC{e_1+e_2+\alpha}{e_6}=+1$\\ 
\xrowht[()]{10pt}
&$\CC{e_1+e_2+\alpha}{S}=\CC{e_1+e_2+\alpha}{b_1+z_1+\alpha}=\CC{e_1+e_2+\alpha}{\tilde{x}+z_1+\alpha}=\CC{e_1+e_2+\alpha}{z_2}=+1$\\
\hline
\xrowht[()]{10pt}
$e_1+e_2+z_1+\alpha $ &$\CC{e_1+e_2+z_1+\alpha}{e_3}=\CC{e_1+e_2+z_1+\alpha}{e_4}=\CC{e_1+e_2+z_1+\alpha}{e_5}=\CC{e_1+e_2+z_1+\alpha}{e_6}=+1$\\ 
\xrowht[()]{10pt}
&$\CC{e_1+e_2+z_1+\alpha}{S}=\CC{e_1+e_2+z_1+\alpha}{b_1+\alpha}=\CC{e_1+e_2+z_1+\alpha}{\tilde{x}+\alpha}=\CC{e_1+e_2+z_1+\alpha}{z_2}=+1$\\
\hline
\xrowht[()]{10pt}
$e_1+e_2+e_3+\alpha $ &$\CC{e_1+e_2+e_3+\alpha}{e_4}=\CC{e_1+e_2+e_3+\alpha}{e_5}=\CC{e_1+e_2+e_3+\alpha}{e_6}=+1$\\ 
\xrowht[()]{10pt}
&$\CC{e_1+e_2+e_3+\alpha}{S}=\CC{e_1+e_2+e_3+\alpha}{\tilde{x}+z_1+\alpha}=\CC{e_1+e_2+e_3+\alpha}{z_2}=+1$\\
\hline
\xrowht[()]{10pt}
$e_1+e_2+e_3+z_1+\alpha $ &$\CC{e_1+e_2+e_3+z_1+\alpha}{e_4}=\CC{e_1+e_2+e_3+z_1+\alpha}{e_5}=\CC{e_1+e_2+e_3+z_1+\alpha}{e_6}=+1$\\ 
\xrowht[()]{10pt}
&$\CC{e_1+e_2+e_3+z_1+\alpha}{S}=\CC{e_1+e_2+e_3+z_1+\alpha}{\tilde{x}+\alpha}=\CC{e_1+e_2+e_3+z_1+\alpha}{z_2}=+1$\\
\hline
\end{tabular}
\caption{\label{NormSal}\emph{Conditions on GGSO coefficients for the survival of spinorial PS level on-shell tachyons in $S$--models. Only $e_1,e_1+e_2,e_1+e_2+e_3$ combinations are detailed but other combinations of $e_i$'s are similar.}}
\end{table}

\FloatBarrier
\subsection{Massless Sectors}
Since the basis is identical to that analysed in \cite{acfkr}, the massless sectors are the 
same, and we will not repeat all the details here. The aspects of the analysis which deserve further
exploration are those in which the absence of supersymmetry manifests itself. Of particular interest
is how the breaking of SUSY by a GGSO phase in $S$--models causes differences to where it is broken
explicitly in the $ \tilde{S}$--models. 

The way of breaking SUSY in the $S$--models (by a GGSO phase) allows the possibility of spontaneous SUSY breaking of
Scherk-Schwarz type \cite{SS, SSS}, 
which cannot be the case for any \ds--models. In particular, it is
well known in orbifold compactifications that if a freely acting orbifold gives a mass contribution
to a gravitino then the SUSY breaking is spontaneous and above the gravitino mass scale SUSY is
effectively restored. In such cases, the orbifold model is an example of a stringy Scherk-Schwarz
compactification. In order to see how this is implemented within our $S$--models, we could transform
it into a orbifold model with freely acting action using the dictionary of \cite{panosdictionary}.
Such an analysis is done explicitly for several $SO(10)$ models in ref. \cite{FR} but doing so for the $S$--models under analysis is left for future work. 

As in the supersymmetric case, spinorial $\mathbf{16}$/$\overline{\mathbf{16}}$'s arise from the sectors
\begin{eqnarray}\label{S16s}
B_{pqrs}^{(1)} &=& S + b_1 + pe_3 + qe_4 + re_5 + se_6
\nonumber \\ 
&=& \{\psi^{\mu},\chi^{1,2},(1-p)y^3\bar{y}^3,
pw^3\bar{w}^3,(1-q)y^4\bar{y}^4,qw^4\bar{w}^4,
 \\
& & ~~~ (1-r)y^5\bar{y}^5,rw^5\bar{w}^5,(1-s)y^6\bar{y}^6,
sw^6\bar{w}^6,\bar{\eta}^{1},\bar{\psi}^{1,\ldots,5}\}
\nonumber \\
B_{pqrs}^{(2)} &=&  S + b_2 + pe_1 + qe_2 + re_5 + se_6\nonumber \\
B_{pqrs}^{(3)} &=&  S + b_3 + pe_1 + qe_2 + re_3 + se_4\nonumber
\end{eqnarray}
where $p,q,r,s = 0,1$ and $b_3 = b_1 + b_2 + x$ and vectorial $\mathbf{10}$'s from the sectors 
\begin{eqnarray}
V_{pqrs}^{(I)} &=&  B_{pqrs}^{(I)}+x\ ,\ I=1,2,3
\label{ovsinovs}
\end{eqnarray}
However, in the SUSY case we know that any states in the Hilbert space of a model will be accompanied by their superpartners, generated through the addition of the $S$ vector. In the non--SUSY case under consideration here, due to the $S$ projection, the would-be superpartners can either be projected or have mismatched charges.  

Since the same formulae can be found for the SUSY classifications we will write the equations for the key classification numbers related to the observable gauge factors without further explanation. From the spinorial $\mathbf{16}/\mathbf{\overline{16}}$ we get the PS numbers
\begin{eqnarray}\label{SpinorialNumbers}
n_{4L} &=& \frac{1}{4}\sum_{\substack{A=1,2,3 \\ p,q,r,s=0,1}} 
P_{pqrs}^A\left(1 + X^A_{pqrs}\right)\left(1 + \CC{B^A_{pqrs}}{\alpha}
\right)\\
n_{4R} &=& \frac{1}{4}\sum_{\substack{A=1,2,3 \\ p,q,r,s=0,1}} 
P_{pqrs}^A\left(1 - X^A_{pqrs}\right)\left(1 - \CC{B^A_{pqrs}}{\alpha}
\right) \\
n_{\overline{4}L} &=& \frac{1}{4}\sum_{\substack{A=1,2,3 \\ p,q,r,s=0,1}} 
P_{pqrs}^A\left(1 - X^A_{pqrs}\right)\left(1 + \CC{B^A_{pqrs}}{\alpha}
\right)\\
n_{\overline{4}R} &=& \frac{1}{4}\sum_{\substack{A=1,2,3 \\ p,q,r,s=0,1}} 
P_{pqrs}^A\left(1 + X^A_{pqrs}\right)\left(1 - \CC{B^A_{pqrs}}{\alpha}
\right).
\end{eqnarray}
where $X^A_{pqrs}$ determines the chirality $\mathbf{16}/\mathbf{\overline{16}}$ of a $B^{1,2,3}_{pqrs}$ sector given by eq. (\ref{S16s}) and $P_{pqrs}^A$ are the projectors (\ref{SpinProjector}) for $B_{pqrs}^{1,2,3}$. Meanwhile, from the $\mathbf{10}$ we can write the bi--doublet and $D(\mathbf{6},\mathbf{1},\mathbf{1})$ numbers as 
\begin{eqnarray}\label{bidoublet}
n_h &=& \frac{1}{2}\sum_{\substack{A=1,2,3 \\ p,q,r,s=0,1 }}R^{(A)}_{pqrs}\left(1 - \CC{V^A_{pqrs}}{\alpha}\right) \\
n_6 &=& \frac{1}{2}\sum_{\substack{A=1,2,3 \\ p,q,r,s=0,1 }}R^{(A)}_{pqrs}\left(1 + \CC{V^A_{pqrs}}{\alpha}\right) 
\end{eqnarray}
where $R^{(A)}_{pqrs}$ are the vectorial projectors in this case. 

The other important aspects of these models that we will need for a phenomenological 
analysis are the absence of gauge enhancements affecting the observable gauge factors and 
the absence of chiral exotic sectors. The exotic sectors listed in Section 2 of the 
SUSY PS classification of \cite{acfkr} are also present in our analysis. However, by 
analysing the  fermionic exotics in our $S$--models we cannot say anything about the 
would-be superpartner scalar exotics. For example, in the SUSY case, if you find an 
absence of any fermionic exotic sectors then a model can be declared exophobic 
(free of all exotic sectors) since the scalar superpartners will also be absent. For our purposes we will not classify these scalar exotic sectors but only 
seek to ensure the absence of chiral exotics by inspecting the fermionic exotics. 

A further difference to our exotics analysis is that we have the exotic sectors 
\beq \label{SExotics}
\begin{Bmatrix}
\{\bar{\lambda^i}\}\ket{S+\alpha}, \ \ \ \ \{\bar{\lambda^i}\}\ket{S+z_1+\alpha} \\
 \ket{S+z_2+\alpha}, \ \ \ \  \ket{S+z_1+z_2+\alpha}\\
 \ket{S+\alpha+x}, \ \ \ \  \ket{S+z_1+\alpha+x} \\
\end{Bmatrix}
\eeq
which in SUSY models are thought of as (gaugino) superpartners of mixed enhancement sectors generating additional gauge bosons. In the SUSY case, such sectors can be assured to be projected out whenever the enhancements are projected. Since the absence of observable enhancements is a fundamental phenomenological constraint to impose, once checked we can be sure that these sectors (\ref{SExotics}) are projected too. For the non--SUSY case, we cannot be sure that the absence of enhancements to the observable gauge group will result in the projection of these exotic states (\ref{SExotics}) so we must account for them when looking at the absence of a chiral exotic anomaly. 
\subsection{Fertility Conditions}\label{SFertConds}
We will use a similar fertility methodology as in the $\tilde{S}$ case except in this case we can consistently enforce conditions on the presence of a Heavy Higgs to break the PS group and have the accompanying Missing Partner Mechanism. We can list the set of fertility conditions as follows: 
\begin{enumerate}
\item[1)] Absence of the tachyonic sectors: 
\beq \label{z1TachCondS}
z_1, \ e_i+z_1, \ e_i+e_j+z_1,\ e_i+e_j+e_k+z_1
\eeq 
using the set of conditions delineated in Table $\ref{NormSSO10}$.
\item[2)] Constraints on $SO(10)$ spinorial states related to the presence of complete fermion families and a Heavy Higgs
\begin{align}\label{Sfertility1}
 n_{4L}-n_{\bar{4}L}  =  n_{\bar{4}R} -n_{4R}\geq 6 \ ,\ n_{\bar{4}R}^{(B)} > 1.
\end{align}

\item[3)] For the presence of a SM higgs, \textit{i.e.} $n_h\geq 1$, and a $D(6,1,1)$, \textit{i.e.} $n_6 \geq 1$, we can impose
\begin{align}\label{Sfertility2}
N_{10}\geq 2
\end{align}
at the $SO(10)$ level. 
\item[4)] Presence of a top quark mass coupling (TQMC) which amounts to fixing the following GGSO coefficients following the methodology of \cite{tqmc}
\begin{eqnarray}\label{Stqmc}
    \CC{e_i}{b_1}&=&-\CC{e_i}{S}, \ \ \  i=1,2 \\
    \CC{e_j}{b_2}&=&-\CC{e_j}{S}, \ \ \ \ j=3,4 \\
    \CC{z_k}{b_1}&=&-\CC{z_k}{S}, \ \ \ \CC{z_k}{b_2}=-\CC{z_k}{S}, \ \ \ \ k=1,2 \\
    \CC{b_1}{e_5}&=&\CC{b_2}{e_5}, \ \ \ \ \CC{b_1}{e_6}=\CC{b_2}{e_6} 
\end{eqnarray}
at the $SO(10)$ level, and
\begin{align} \label{StqmcPS}
\CC{b_1}{\alpha} =-\CC{b_2}{\alpha} = \CC{S}{\alpha}
\end{align}
at the PS level.
\end{enumerate}
A few comments on these conditions are in order. 

In condition 2) the equation $n_{\bar{4}R}^{(B)} > 1$ is a necessary, but not sufficient, condition on the presence of a Heavy Higgs to break the PS gauge symmetry. We specify that these come from the bosonic sectors $B^{(1,2,3)B}_{pqrs}=S+B^{(1,2,3)F}_{pqrs}$. We note that any such sector, $\xi$, that survives the GGSO projections and has a GGSO phase $\CC{\xi}{\alpha}=1$ will give rise to the scalars with observable representations
\beq 
\left[ \binom{3}{\text{even}}_{\bar{\psi}^{1,2,3}}+\binom{3}{\text{odd}}_{\bar{\psi}^{1,2,3}}\right]\binom{2}{\text{even}}_{\bar{\psi}^{4,5}} =: n_{\bar{4}R}^{(B)}+n_{4R}^{(B)},
\eeq 
where we employ the convenient notation $\binom{3}{even}=\{ \ket{\pm \pm \pm} \ | \ \# (\ket{-}) \mod  2=0\}$. These $n_{\bar{4}R}^{(B)}+n_{4R}^{(B)}$ states are those we want to break the PS gauge group. Furthermore, with both of them 
we can allow for breaking along a D-flat direction, which enables hierarchical SUSY breaking via Scherk-Schwarz, although this would need to be evaluated carefully for a specific model. 

In condition 3) imposing $n_6 \geq 1$ is required for the implementation of the Missing Partner Mechanism of PS models. 
This occurs once the PS group is broken and the Heavy Higgs has acquired a VEV. With a $(\mathbf{6},\mathbf{1},\mathbf{1})$ triplet/anti--triplet field we induce couplings to $d/\bar{d}$ SM fields of the form $\sim \langle n_{4R} \rangle d^cD_3 $ and 
$\sim \langle n_{4R} \rangle \bar{d}^c\bar{D}_3 $. These form massive states at the GUT scale which protect proton decay from happening too quickly.

\section{Partition Function of Pati--Salam Models}\label{PF}

The analysis of the partition function is particularly important in non--supersymmetric constructions, as it gives a complementary tool to analyse the structure of the theory. It also provides the necessary tools to count the total number of states at each mass level and hence check for the existence of on and off--shell tachyons in any specific model. Moreover, its integration over the fundamental domain corresponds to the cosmological constant. 

The partition function for free fermionic theories is given by the integral
\begin{equation}
    Z = \int_\mathcal{F}\frac{d^2\tau}{\tau_2^2}\, Z_B  \sum_{\alpha,\beta} \CC{\alpha}{\beta} \prod_{f} Z \sqbinom{\alpha(f)}{\beta(f)},
    \label{ZInt}
\end{equation}
where $d^2\tau/\tau_2^2$ is the modular invariant measure and $Z_B$ 
is the bosonic contribution arising from the worldsheet bosons.  The sum gives the contributions from the worldsheet fermions and involves all combinations of basis vectors $\alpha,\beta$ from a given basis set, which in our case are the ones given in (\ref{basisSt}) and (\ref{basisS}).  The integral is over the fundamental domain of the modular group 
$$ \mathcal{F} = \{\tau\in\mathbb{C}\,|\,|\tau|^2>1 \;\land\;|\tau_1|<1/2\}, $$
which ensures that only physically inequivalent geometries are counted. The above integral specifically represents the one--loop vacuum energy $\Lambda$ of our theory. Note that this is the cosmological constant from the worldsheet point of view and hence is a dimensionless quantity \cite{DienesLandscape}. It is related to the spacetime cosmological constant, $\lambda$, by $\lambda =  -\frac{1}{2}\mathcal{M}^4\Lambda$, where $\mathcal{M}=M_{String}/2\pi$. For simplicity, in the following we will refer to $\Lambda$ as the cosmological constant.

The best way to perform this integral is as presented in \cite{so10tclass,D} using the expansion of the $\eta$ and $\theta$ functions in terms of the modular parameter $\tau$, or more more precisely, in terms of $q\equiv e^{2\pi i \tau}$ and $\bar{q}\equiv e^{-2\pi i \bar{\tau}}$. This leads to a series expansion of the one-loop partition function 
\begin{equation}
    Z = \sum_{n.m} a_{mn} \int_\mathcal{F} \frac{d^2\tau}{\tau_2^3} \, q^m \bar{q}^n.
    \label{QPF}
\end{equation}
The benefit of such an expansion to the partition function is that the $a_{mn}$ 
represent the difference between bosonic and fermionic degrees of freedom at each mass 
level, i.e. $a_{mn} = N_b - N_f$.  
As expected, on--shell tachyonic states, {\it i.e.} states 
with $m=n<0$, have an infinite contribution. On the other hand off--shell tachyonic states 
may contribute a finite value to the partition function. It is indeed known that such 
off--shell tachyonic states are necessarily present in the spectrum of non--SUSY 
theories \cite{D} and so all of our Pati--Salam models posses such states. 
It is also important to note that modular invariance constraints imposed on the 
basis and GGSO phases only allows states with $m-n\in \mathbb{Z}$.

In theories with spacetime supersymmetry, it is ensured that the bosonic and fermionic degrees of
freedom are exactly matched at each mass level. That is, we necessarily have that $a_{mn} = 0$ for
all $m$ and $n$, which in turn causes the vanishing of the cosmological constant. For
non--supersymmetric models like the ones introduced in Sections \ref{StPS} and \ref{PSNormS}, this
level-by-level cancellation is not ensured and so in general produce a non--zero value for $\Lambda$.
This value, however, is specific to the fermionic point in the orbifold moduli space and hence moving
away from this point can in principle change $\Lambda$ as discussed in \cite{FR}. 

An interesting possibility is to try to suppress the value of the cosmological constant so that we
are guaranteed a small value. It has been shown in \cite{NoScaleItoyama1, NoScaleItoyama2, ADMNSUSY,
NoScaleKP2, NoScaleKP1} that such a mechanism may be possible for models which satisfy the constraint
$N^0_b=N^0_f$ or in the above language $a_{00} = 0$. These rely on a Scherk--Schwarz SUSY breaking in
which the internal dimensions of the compactification space $R$ are used to suppress the next--to
leading order contributions to $\Lambda$ in the large volume $R\to\infty$ limit. This is because in
this setup, the dependence of the cosmological constant on the compactification radius schematically
becomes
\begin{equation}
\Lambda \propto (N^0_b - N^0_f) \frac{1}{R^4} + \mathcal{O}(e^{-\alpha R^2}),
\end{equation}
and thus in the large $R$ limit, $\Lambda$ is exponentially suppressed given that $N_b^0 = N_f^0$.
Thus in our classification program, we also try to find models which fulfil this condition and so may
be good candidates to further explore this idea. It is important to note, however, that since the
above mechanism relies on a Scherk-Schwarz breaking, it can only be applied to (some) $S$--models.
This is because in $\tilde{S}$--models, SUSY is broken explicitly.

\section{Results of Classification} \label{results}
Using the methodology built-up in the previous sections, we can now turn towards analysing samples of PS $\tilde{S}$ and $S$--models with respect to standard phenomenological criteria. First of all, we perform a random classification in the space of PS $\tilde{S}$ and $S$--models. This works by generating random GGSO phase configurations and classifying according to the absence of tachyons and the classification numbers: $n_{4L},n_{\bar{4}L},n_{4R},n_{\bar{4}R}, n_h, n_6, n_{4},n_{\bar{4}},$ $n_{2L},n_{2R}$ which are common to $\tilde{S}$ and $S$--models and are defined in previous sections. For the $\tilde{S}$--models there are a further two classification numbers, $n_{V2L},n_{V2R}$, relating to the vectorial exotics (\ref{ExotVecs}). The results of the random classification for a sample of $10^9$ GGSO configurations in both the $\tilde{S}$ and $S$--model cases are displayed in Table \ref{RandomStStats} and \ref{RandomSStats}, respectively. 
\begin{table}[!htb]
\centering
\begin{tabular}{|c|l|r|c|r|}
\hline
&Constraints & \parbox[c]{2.5cm}{Total models in sample}& Probability \\
\hline
 & No Constraints & $ 10^9$ & $1$  \\ \hline
(1)&{+ Tachyon--free} & 2038657 & $2.04\times 10^{-3}$  \\  \hline
(2)& {+ No Observable Enhancements} & 2014917 & $2.01\times 10^{-3}$  \\ \hline
(3)&{+ Complete Families} & 572411 & $5.72\times 10^{-4}$  \\  \hline
(4)&{+ No Chiral Exotics} & 403989 & $4.04\times 10^{-4}$  \\  \hline
(5)&{+ 3 Generations}& 3074 & $3.07\times 10^{-6}$ \\ \hline
(6)&{+ $n_{\bar{4}R}^F,n_{4R}^F$ Present}& 346& $3.46 \times 10^{-7}$ \\ \hline
(7)&{+ SM Higgs}& 314&$3.14\times 10^{-7}$ \\ \hline
(8)&{+ Heavy Triplet Constraint}& 298&$2.98\times 10^{-7}$ \\ \hline
(9)&{+ TQMC }& 289 & $2.89\times 10^{-7}$ \\ \hline
(10)&{+ No Fermionic Exotics}& $0$ & 0\\ \hline
\end{tabular}
\caption{\label{RandomStStats} \emph{Phenomenological statistics from a sample of $ 10^9$ randomly generated Pati--Salam $\tilde{S}$--models.}}
\end{table}
\begin{table}[!htb]
\centering
\begin{tabular}{|c|l|r|c|r|}
\hline
&Constraints & \parbox[c]{2.5cm}{Total models in sample}& Probability \\
\hline
 & No Constraints & $ 10^9$ & $1$  \\ \hline
(1)&{+ Tachyon--free} & 10578258 & $1.06\times 10^{-2}$  \\  \hline
(2)& {+ No Observable Enhancements} & 10246688 & $1.02\times 10^{-2}$  \\ \hline
(3)&{+ Complete Families} & 2730363 & $2.73\times 10^{-3}$  \\  \hline
(4)&{+ No Chiral Exotics} & 461666 & $4.62\times 10^{-4}$  \\  \hline
(5)&{+ 3 Generations}& 3103 & $3.10\times 10^{-6}$ \\ \hline
(6)&{+ PS Higgs}& 2684&$2.68\times 10^{-7}$ \\ \hline
(7)&{+ SM Higgs}& 2263&$2.26\times 10^{-7}$ \\ \hline
(8)&{+ Partner Mechanism}& 1934&$1.93\times 10^{-7}$ \\ \hline
(9)&{+ TQMC}& 1878 & $1.88\times 10^{-7}$\\ \hline
(10)&{+ No Fermionic Exotics}& 36& $3.6\times 10^{-9}$ \\ \hline
\end{tabular}
\caption{\label{RandomSStats} \emph{Phenomenological statistics from a sample of $ 10^9$ randomly generated Pati--Salam $S$--models.}}
\end{table}
\FloatBarrier
From these results we can first note a higher probability of $\sim 0.01$ that an $S$--model is tachyon--free compared with an $\tilde{S}$--model which has probability $\sim 0.002$. This is likely due to the power of GGSO phases involving $S$ in projecting the tachyons. Another notable result is the absence of PS $\tilde{S}$--models without massless fermionic exotics in the final criteria (10), whereas we find vacua for the $S$--models where all fermionic exotic sectors are projected. As mentioned in previous sections, we are not classifying the scalar exotic sectors here and so we can not say whether there are exophobic models, which were found in the SUSY PS classification of \cite{acfkr}. Due to the absence of SUSY it appears that finding such exophobic models would be much less likely due to having to check the bosonic exotic sectors independently to the fermionic ones. 

As explained in earlier sections, the PS \ds--model analysis is unphysical in the sense that we implement the condition (6) for the presence of $n_{\bar{4}R}^F,n_{4R}^F$ and condition (8) for the presence of a $D(\mathbf{6},\mathbf{1},\mathbf{1})$ in order to draw an analogy with the analysis of the $S$--models where a Heavy Higgs can be found from the bosonic component of the $\mathbf{16}/\mathbf{\overline{16}}$ and a Missing Partner Mechanism can be implemented requiring the $D(\mathbf{6},\mathbf{1},\mathbf{1})$ state. The difference in condition (6) skew the numbers a bit since requiring a PS Higgs for the $S$--models is a weak condition, whereas the symbolic condition for a $n_{\bar{4}R}^F,n_{4R}^F$ in the \ds--models is stronger.

In order to find more models satisfying our phenomenological constraints we can now turn to utilizing the fertility methodology outlined in sections \ref{StFertConds} and \ref{SFertConds} for the $\tilde{S}$ and $S$ cases, respectively. Explicitly, we do this by implementing the conditions 1) - 4) in sections \ref{StFertConds} and \ref{SFertConds} in the space of $SO(10)$--models, $\Pi_{SO(10)}$, and collect 200,000 fertile $\tilde{S}$ $SO(10)$ cores and 550,000 $S$ cores. Once we have the cores, we look at the space $\Pi_{\alpha}$ and, in particular, we iterate over all independent choices of $\alpha$ GGSO phases. When equations (\ref{tqmcPS}) and (\ref{StqmcPS}) are accounted for we are left with 10 independent such phases. However, a subtlety arises here due to how to avoid SUSY configurations in the $S$--models. We must allow for the possibility that some $SO(10)$ cores are supersymmetric in the $S$--models because we can still get non-SUSY models in the case where SUSY is broken by $\alpha$, \textit{i.e.} $\CC{S}{\alpha}=+1$. The logical procedure to deal with this is to check if the condition (\ref{SUSYphases}) holds for the $SO(10)$ phases and then fix $\CC{S}{\alpha}=+1$ and iterate over the reduced $\Pi_{\alpha}$ space spanned by 9 independent GGSO phases involving $\alpha$. In our sample of $S$ $SO(10)$ cores we extracted 50,000 such SUSY cores and 500,000 non-SUSY cores

The results of this fertility classification are displayed in Table \ref{FertileStStats} and \ref{FertileSStats}.
\begin{table}[!htb]
\centering
\begin{tabular}{|c|l|r|c|r|}
\hline
&Constraints & \parbox[c]{2.5cm}{Total models in sample}& Probability \\
\hline
 & No Constraints & $ 204800000$ & $1$  \\ \hline
(1)&{+ Tachyon--free} &  $2417463$ & $1.18\times 10^{-2}$\\  \hline
(2)& {+ No Observable Enhancements} &  $2406298$ & $1.17\times 10^{-2}$\\ \hline
(3)&{+ Complete Families} &  $623004$ & $3.04\times 10^{-3}$\\  \hline
(4)&{+ No Chiral Exotics} &  $438280$ & $2.14\times 10^{-3}$\\  \hline
(5)&{+ 3 Generations}&  $327463$ & $1.60\times 10^{-3}$\\ \hline
(6)&{+ PS Higgs}& $190766$ &$9.31\times 10^{-4}$\\ \hline
(7)&{+ SM Higgs}& $190766$ &$9.31\times 10^{-4}$\\ \hline
(8)&{+ Partner Mechanism}& $183753$ & $8.97\times 10^{-4}$\\ \hline
(9)&{+ TQMC }& $183753$ & $8.97\times 10^{-4}$\\ \hline
(10)&{+ No Fermionic Exotics}& $0$ &0\\ \hline
\end{tabular}
\caption{\label{FertileStStats} \emph{Phenomenological statistics for  Pati--Salam $\tilde{S}$--models derived from 200000 fertile $SO(10)$ $\tilde{S}$--cores.}}
\end{table}
\begin{table}[!htb]
\centering
\begin{tabular}{|c|l|r|c|r|}
\hline
&Constraints & \parbox[c]{2.5cm}{Total models in sample}& Probability \\
\hline
 & No Constraints &   537600000 & $1$  \\ \hline
(1)&{+ Tachyon--free} &  11770044 & $2.19\times 10^{-2}$\\  \hline
(2)& {+ No Observable Enhancements} &  11431950 & $2.12\times 10^{-2}$ \\ \hline
(3)&{+ Complete Families} &  3020242 & $5.62\times 10^{-3}$\\  \hline
(4)&{+ No Chiral Exotics} &  723352 & $1.35\times 10^{-3}$ \\  \hline
(5)&{+ 3 Generations}&  488802 & $9.09\times 10^{-4}$ \\ \hline
(6)&{+ PS Higgs}& 444454 &$8.27\times 10^{-4}$\\ \hline
(7)&{+ SM Higgs}& 444454 &$8.27\times 10^{-4}$\\ \hline
(8)&{+ Partner Mechanism}& 384080 &$7.14\times 10^{-4}$ \\ \hline
(9)&{+ TQMC}& 384080 &$7.14\times 10^{-4}$ \\ \hline
(10)&{+ No Fermionic Exotics}& 16030 &$2.98\times 10^{-5}$\\ \hline
\end{tabular}
\caption{\label{FertileSStats} \emph{Phenomenological statistics for 537600000 PS $S$--models derived from 550000 fertile $SO(10)$ S--cores.}}
\end{table}
\FloatBarrier
The first thing to note from these results is that owing to the fertility analysis the number of models meeting the phenomenological criteria has improved by around 3 orders of magnitude compared with the random classification. Furthermore, as expected, we see that all PS models derived in the fertility methodology will 
come with a SM Higgs and TQMC automatically. 
The final classification data we will display can be found in Appendix A, where the quantum numbers for models satisfying constraints (1)-(9) in Tables \ref{RandomSStats} and \ref{FertileSStats} are displayed. 
\subsection{Results for $N_b^0-N_f^0$}
\begin{figure}[!htb]
\centering
\includegraphics[width=0.85\linewidth]{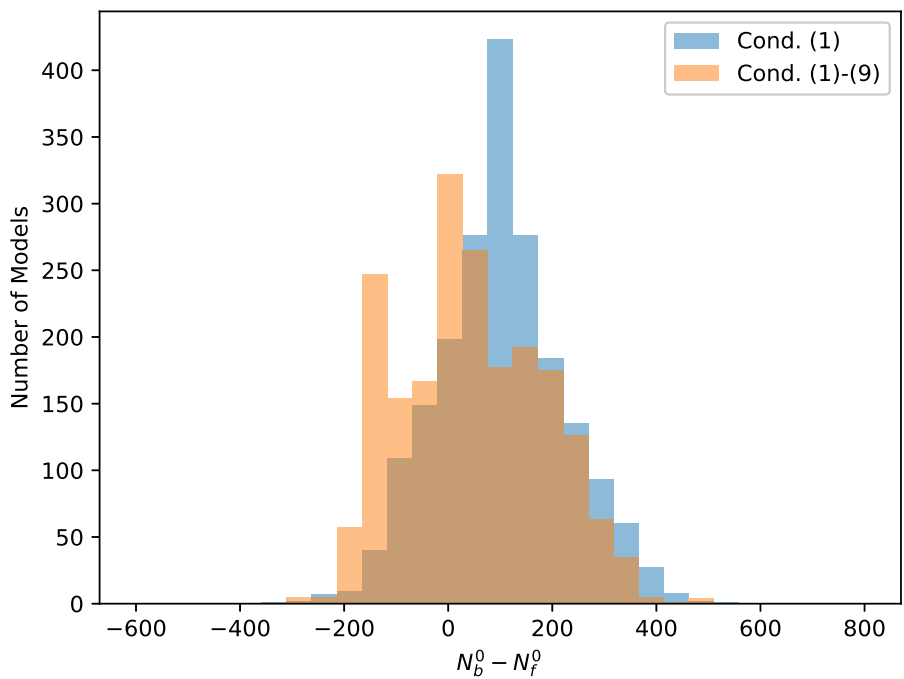}
\caption{\emph{The distribution of the constant term $a_{00} = N_b^0-N_f^0$ for a sample of $2\times10^3$ $\tilde{S}$--models satisfying conditions (1) and (1)-(9) of Table \ref{FertileStStats} .}}
\label{STDist}
\end{figure}

\begin{figure}[!htb]
\centering
\includegraphics[width=0.85\linewidth]{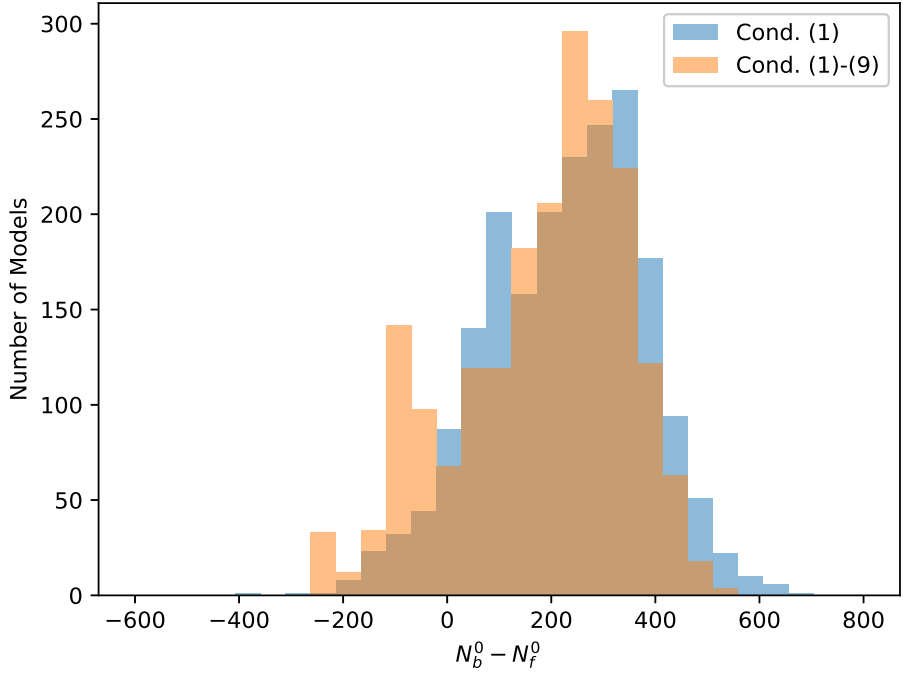}
\caption{\emph{The distribution of the constant term $a_{00} = N_b^0-N_f^0$ for a sample of $2\times10^3$ $S$--models satisfying conditions (1) and (1)-(9) of Table \ref{FertileSStats} .}}
\label{SDist}
\end{figure}

As discussed in Section \ref{PF}, the constant term of the partition function, $a_{00} = N_b^0-N_f^0$, is an important quantity. It quantifies the Bose--Fermi degeneracy at the massless level and so is of phenomenological significance. It also provides the leading order behaviour of the vacuum energy, and thus models with $N_b^0-N_f^0$ can be of particular interest. Thus, in our classification program we have taken a close look at this value. Its distribution for a sample of $2\times 10^3$ is shown in Figures \ref{STDist} and \ref{SDist} for $\tilde{S}$ and $S$--models respectively. We have included a random sample of non--tachyonic vacua along with a sample of models satisfying the criteria (1)-(9) in order to see what effect the imposition of certain phenomenological features have on this net Bose--Fermi degeneracy. The distributions show that for both sets of models it should be possible to find ones which satisfy the condition $N^0_b=N^0_f$. Indeed, from the sample of $2\times 10^3$ we have found 14 such models for both $\tilde{S}$ and $S$.

\FloatBarrier
\subsection{Example Models with $N_b^0=N_f^0$}
Having found $\mathcal{O}(10^5)$  $\tilde{S}$ models satisfying all phenomenological constraints (1)-(9) we can do a subsequent search for such models that additionally satisfy the condition $a_{00}=N^0_b-N^0_f=0$ which suppresses the leading order contribution to the cosmological constant of the model. The following GGSO phase configuration meets this condition
\begin{equation}
\CC{v_i}{v_j}= 
\begin{blockarray}{rrrrrrrrrrrrrr}
&\mathds{1}& \tilde{S} & e_1 & e_2 & e_3 & e_4 & e_5 & e_6 & b_1 & b_2 & b_3 & z_1 & \alpha \\
\begin{block}{r(rrrrrrrrrrrrr)}
 \mathds{1}&-1&  1&  1&  1& -1& -1& -1& -1&  1& -1&  1&  1& -1\ \\ 
 \tilde{S}& 1& -1&  1& -1&  1& -1&  1&  1&  1& -1& -1&  1& -1\ \\ 
 e_1& 1&  1& -1& -1& -1&  1&  1&  1& -1& -1& -1& -1&  1\ \\ 
 e_2& 1& -1& -1& -1& -1&  1& -1& -1& -1& -1& -1& -1&  1\ \\ 
 e_3& -1&  1& -1& -1&  1&  1&  1&  1& -1& -1& -1& -1& -1\ \\ 
 e_4& -1& -1&  1&  1&  1&  1&  1& -1& -1& -1& -1& -1&  1\ \\ 
 e_5& -1&  1&  1& -1&  1&  1&  1& -1& -1& -1& -1&  1&  1\ \\ 
 e_6& -1&  1&  1& -1&  1& -1& -1&  1& -1& -1& -1& -1&  1\ \\ 
 b_1&1& -1& -1& -1& -1& -1& -1& -1&  1& -1& -1& -1&  1\ \\ 
 b_2&-1&  1& -1& -1& -1& -1& -1& -1& -1& -1& -1& -1& -1\ \\ 
 b_3&1&  1& -1& -1& -1& -1& -1& -1& -1& -1&  1& -1&  1\ \\ 
 z_1&1& -1& -1& -1& -1& -1&  1& -1& -1& -1& -1&  1&  1\ \\ 
 \alpha&-1& -1&  1&  1& -1&  1&  1&  1& -1&  1& -1& -1& -1\ \\
 \end{block}
\end{blockarray}
\end{equation}
Furthermore, as desired, it enjoys a PS--breaking Higgs and 3 chiral generations with PS quantum numbers $n_{4L}=3, n_{\bar{4}L}=0, n_{4R}=1$ and  $n_{\bar{4}R}=4$. Furthermore, the model has the necessary SM Higgs and $D(6,1,1)$ for the unphysical `Heavy Triplet Constraint', such that $n_h=3$ and $n_6=5$. The model also has exotic quantum numbers $n_{4}=n_{\bar{4}}=1,n_{2L}=n_{2R}=6$ and $n_{V2L}=n_{V2R}=0$ which allows for the generation of vector--like exotics at high mass scale so as not to have fractionally charged states at lower mass scales which violates experimental observation. The top quark mass coupling is guaranteed by the conditions (\ref{Stqmc}) and (\ref{StqmcPS}). We also note that the $\tilde{x}$--sector arises in the spectrum of this model. This is important since it is charged under the observable
group and in this case generates four extra $n_{\bar{4}R}$ and $n_{4R}$ charged under each $U_{1,2,3}$ factor, leaving the number of generations still equal to 3. 
It was noted in \cite{so10tclass} that the $\tilde{x}$--sector corresponds to the sector producing the
fermionic superpartners of the states from the $x$--sector, {\it i.e.}
$S+x$, which enhance the $SO(10)$ symmetry to $E_6$. The $\tilde{x}$--sector
therefore gives rise to the fermionic superpartners of the spacetime
vector bosons from the $x$--sector, which do not arise in the construction of our $\tilde{S}$--models. 
We can also calculate the traces under the $U(1)_{i=1,2,3}$ associated with the 
right--moving currents $\bar{\eta}^i\bar{\eta}^i$, which are
\beq 
 \mathrm{Tr}\ U(1)_1=-24, \ \ \ \  \mathrm{Tr}\ U(1)_2=-24 \ \text{ and } \  \mathrm{Tr}\ U(1)_3=48.
\eeq 
such that the combination $U(1)_1+U(1)_2-U(1)_3$ is anomalous and we can choose $U(1)_1-U(1)_2$ and $U(1)_1+U(1)_2+U(1)_3$ as anomaly-free combinations.
The partition function for this model is given by
\beq\label{PFStModel}
Z = 2\,q^0\bar{q}^{-1} + 0\,q^0\bar{q}^0 - 668\,q^{1/8}\bar{q}^{1/8} - 4224\,q^{1/4}\bar{q}^{1/4} + 32\,q^{3/8}\bar{q}^{-5/8} + \cdots ,
\eeq
where we see that there are no on--shell tachyons and that we have equal number of 
bosons and fermions at the massless level as advertised. We can further note the off--shell
model--independent term $2\,q^0\bar{q}^{-1}$ obtained from the so--called `proto--graviton' resulting
from the state $\psi^\mu \ket{0}_L\otimes \ket{0}_R$. This expression (\ref{PFStModel}) integrates
via (\ref{QPF}) to give a cosmological constant 
$$ \Lambda=-166.42.$$
Recall that, as described in Section \ref{PF}, this is the dimensionless worldsheet vacuum energy and
hence has the opposite sign compared to the 4D spacetime cosmological constant. 

Turning our attention to the analysis of $S$--models, we can achieve $N^0_b=N^0_f$ and meet the phenomenological constraints (1)-(9) with the following GGSO phase configuration 
\begin{equation}
\label{GGSOphasesS}
\CC{v_i}{v_j}= 
\begin{blockarray}{rrrrrrrrrrrrrr}
&\mathds{1}& S & e_1 & e_2 & e_3 & e_4 & e_5 & e_6 & b_1 & b_2 & z_1 & z_2 & \alpha \\
\begin{block}{r(rrrrrrrrrrrrr)}
 \mathds{1}&-1& -1& -1&  1& -1&  1& -1& -1& -1&  1&  1&  1& -1\ \\ 
 S&-1& -1& -1& -1& -1& -1& -1& -1& -1&  1&  1& -1& -1\ \\ 
 e_1&-1& -1&  1& -1&  1&  1&  1& -1&  1&  1&  1&  1&  1\ \\ 
 e_2& 1& -1& -1& -1&  1&  1&  1&  1&  1& -1&  1&  1& -1\ \\ 
 e_3&-1& -1&  1&  1&  1&  1&  1& -1& -1&  1& -1&  1&  1\ \\ 
 e_4& 1& -1&  1&  1&  1& -1&  1&  1& -1&  1&  1&  1& -1\ \\ 
 e_5&-1& -1&  1&  1&  1&  1&  1& -1& -1& -1&  1& -1&  1\ \\ 
 e_6&-1& -1& -1&  1& -1&  1& -1&  1&  1&  1& -1&  1& -1\ \\ 
 b_1&-1&  1&  1&  1& -1& -1& -1&  1& -1&  1& -1&  1& -1\ \\ 
 b_2& 1& -1&  1& -1&  1&  1& -1&  1&  1&  1& -1&  1&  1\ \\ 
 z_1& 1&  1&  1&  1& -1&  1&  1& -1& -1& -1&  1&  1&  1\ \\ 
 z_2& 1& -1&  1&  1&  1&  1& -1&  1&  1&  1&  1&  1& -1\ \\ 
 \alpha&-1& -1&  1& -1&  1& -1&  1& -1&  1& -1& -1& -1& -1\ \\ 
\end{block}
\end{blockarray}
\end{equation}
which enjoys a PS--breaking Higgs with $\#(n_{4R}/n_{\bar{4}R})=3$
 and three chiral generations with quantum numbers $n_{4L}=4, n_{\bar{4}L}=1, n_{4R}=1$ and $n_{\bar{4}R}=4$. Furthermore the model has the necessary SM Higgs and $D(6,1,1)$ for the Missing Partner Mechanism such that $n_h=2, n_6=2$. The model also has exotic quantum numbers $n_{4}=n_{\bar{4}}=0 $ and $n_{2L}=n_{2R}=4$, which ensures the absence of chiral exotic states. We note that exophobic vacua with $N_b^0=N_f^0$ were not found in our analysis. 
The top quark mass coupling is guaranteed by the conditions (\ref{Stqmc}) and (\ref{StqmcPS}). The traces of $U(1)_{1,2,3}$ for this model are given by:
\beq 
\mathrm{Tr} \ U(1)_1=0, \ \ \ \  \mathrm{Tr}\  U(1)_2=0, \ \text{ and } \  \mathrm{Tr}\ U(1)_3=0
\eeq 
such that each of the $U(1)_{1,2,3}$ are anomaly-free independently. For $U(1)_1$ this cancellation occurs between the trace in the observable $\mathbf{16}/\overline{\mathbf{16}}$ and the trace from hidden sectors. For the $U(1)_2$ and $U(1)_3$ the cancellation happens in each type of sectors (observables, hidden, exotics)  independently. 

Inspecting the GGSO phase matrix (\ref{GGSOphasesS}) we see that supersymmetry is only broken by one phase $\CC{S}{z_1}=1$. It is not too surprising that configurations close to supersymmetric ones are common origins of potentially viable models since they preserve some of the benefits from having supersymmetry. In particular, having most of the GGSO phases involving $S$ equal to $-1$ will help to ensure the absence of tachyons. The gravitino is of course projected but we note that the following states from the $S$ sector: $\{\bar{\psi}^{1,2,3},\bar{\eta}^{1,2,3}\}\{\bar{\phi}^{3,4}\}\ket{S}$ and $\{\bar{\psi}^{4,5}\}\{\bar{\phi}^{1,2}\}\ket{S}$ are retained. 

The partition function for this $S$ model is given by
\beq
Z = 2q^0\bar{q}^{-1} + 0\,q^0\bar{q}^0 + 16\,q^{1/8}\bar{q}^{1/8} - 192\,q^{1/4}\bar{q}^{1/4}  +192\,q^{3/8}\bar{q}^{3/8} - 4\,q^{1/2}\bar{q}^{-1/2} + \cdots,
\eeq
resulting in a worldsheet vacuum energy
$$ \Lambda=-62.66.$$
We see that we indeed have $N^0_b = N^0_f$, hence the lack of constant
term in both models above. We also observe the necessary off--shell
tachyon at $\bar{q}^{-1}$ and the lack of physical tachyons. 

\section{Conclusions}\label{conclude}

The free fermionic representation of the 
heterotic string in four dimensions gave rise to an abundance of three
generation models with
varying unbroken $SO(10)$ subgroups and the canonical GUT
embedding of the weak hypercharge. 
These models correspond to $\mathbb{Z}_2\times\mathbb{Z}_2$ orbifold of
six--dimensional compactified tori
at special points in the moduli space \cite{z2xz2, panosdictionary}. 
The free fermionic formalism was used 
to develop a systematic classification of the
$\mathbb{Z}_2\times\mathbb{Z}_2$ toroidal orbifolds, leading to
numerous fundamental observations, among them: the
construction of the first known string models
that produce in the low energy effective field theory
solely the spectrum of the 
Minimal Supersymmetric Standard Model \cite{slm};
the discovery of spinor--vector duality
in the space of $\mathbb{Z}_2\times\mathbb{Z}_2$ orbifold
compactifications \cite{fkr, cfkr}; 
the discovery of exophobic string models \cite{acfkr}. 

In this paper we extended the systematic classification of free fermionic 
$\mathbb{Z}_2\times\mathbb{Z}_2$ orbifolds to non--supersymmetric
Pati--Salam models. We pursued
the construction of such models via two routes,
based on the \ds--models and $S$--models, 
where the first class descend from a tachyonic ten--dimensional vacuum,
whereas the second
correspond to compactifications of the ten--dimensional non--supersymmetric 
$SO(16)\times SO(16)$ heterotic string. 
A first task in the construction of non--supersymmetric
models is to ensure that all the physical tachyonic states
are projected out from the physical spectrum. 
Systematic classification rules were developed to analyse the tachyon
producing sectors and to extract tachyon--free vacua in the two classes of
models. 
tachyon--free models were found with probability 0.002 and 0.01 in \ds~
and $S$--models, 
respectively. To facilitate the extraction of phenomenological vacua merits the 
development of the `fertility methodology' \cite{slmclass,ferlrs}
that pre--selects
$SO(10)$ preserving GGSO configuration that are amenable for producing viable
phenomenological models. We demonstrated that application of the `fertility 
methodology' increases the frequency of models that satisfy key
phenomenological 
criteria by three orders of magnitude. We note that whereas $S$--models
contain the heavy Higgs scalar representations required to break the 
Pati--Salam gauge symmetry down to the Standard Model gauge group, 
they are in fact absent in Pati--Salam \ds--models. Construction 
of \ds--models that satisfy this criteria is only possible by breaking 
the $SO(10)$ gauge symmetry to the Standard Model subgroup directly at
the string scale. This follows from the fact that the heavy Higgs scalar
representations are also absent in $SU(5)\times U(1)$ \ds--models, as well as 
in those with $SU(3)\times U(1)\times SO(4)$ unbroken $SO(10)$ subgroup, 
whereas the Standard--like models utilise Standard Model singlet
states that are obtained from exotic sectors, as shown in ref. \cite{stable}.  
Additionally, we analysed the partition function several exemplary models 
and demonstrated the existence of three generation models that satisfy the 
desired criteria $a_{00}=N_b^0-N_f^0=0$, {\it i.e.} with equal number of bosonic
and fermionic massless degrees of freed
om. The fermionic $\mathbb{Z}_2\times\mathbb{Z}_2$ orbifolds 
provide the tools to develop the phenomenological approach to string theory. 
With an abundance of models and tools to explore this space of string vacua, 
the stage is now ripe to explore the larger space of unviable constructions, 
{\it \`a la} ref. \cite{type0} and the dynamics that may lie behind the 
string vacuum selection.

We would like also to comment on several issues that warrants further analysis
in non--supersymmetric string configurations. It is well known that
in the Minimal Supersymmetric Standard Model (MSSM) the gauge
couplings of the Standard Model
tend to merge together at a scale of the
order of the GUT scale, whereas they do not if the
spectrum consist solely of the Standard Model states \cite{gcu}.
While the naive MSSM realisation of supersymmetry is under increasing strain
from experimental observations, it may well exist at scales beyond
those of contemporary experiments, and serve to mitigate the
unification of the couplings at the string scale. In this respect we
note that the picture in non--supersymmetric string vacua is more complex.
In the first place, many observable sectors in the string models
may still exhibit Bose--Fermi degeneracy of the Standard Model
states, while differing in their charges under some other symmetries
of the models \cite{aafs}.
Because supersymmetry is broken, these states can no longer
sit in super--multiplets, but as far as the Standard Model charges
are concerned they appear just as in the supersymmetric models.
Furthermore, there may be intermediate states between the string
and electroweak scale that modify the naive picture of the MSSM,
resulting in agreement of the string scale coupling unification
with the low energy data \cite{gcustring}.
The issue of gauge coupling unification can therefore only be examined
on a model by model basis and cannot be addressed in a broad classification,
such as the one that we presented here. 

Another problem of interest in non-supersymmetric string vacua
is the existence of large tadpole diagrams that are
generated due to the non--vanishing vacuum energy and
reflect the instability of the string vacuum. As we
cautioned above, in our view any argument of stability in
non--supersymmetric string vacua is at best speculative.
Nevertheless, we can propose a possibility of how the
issue of tadpole diagrams might be addressed in the class
of models under investigation here. As is well known,
heterotic--string models often contain an anomalous $U(1)_A$ symmetry.
This anomalous $U(1)_A$ generates a non--vanishing amplitude
at one--loop order in string perturbation theory,
and one can envision that it can be used to cancel the
one--loop diagram arising from the non--vanishing vacuum energy.
While this can be expected in general, its realisation
can only be implemented in case by case basis,
as is the case in supersymmetric vacua.
There are subtleties associated with
the calculation in non--supersymmetric backgrounds,
that we hope to return to in a future publication.

\section*{Acknowledgments}

The work of VGM is supported in part by EPSRC grant EP/R513271/1.
The work of BP is supported in part by STFC grant ST/N504130/1.

\pagebreak
\appendix
\section{Key Classification Number Tables}

\begin{table}[!htb]
\setlength{\tabcolsep}{5pt}
\renewcommand{\arraystretch}{0.85}
\scriptsize
\centering
\begin{tabular}{|c|c|c|c|c|c|c|c|c|c|c|}
\hline
$n_{4L}$&$n_{\bar{4}L}$&$n_{4R}$&$n_{\bar{4}R}$&$n_h$&$n_6$&No Fermionic Exotics&Frequency\\ \hline
3& 0& 1& 4& 5& 3& False&    50988\\ \hline
4& 1& 1& 4& 5& 5& False&    27602\\ \hline
3& 0& 1& 4& 3& 5& False&    25090\\ \hline
3& 0& 1& 4& 3& 1& False&    12526\\ \hline
3& 0& 1& 4& 3& 3& False&    10259\\ \hline
4& 1& 1& 4& 3& 3& False&     8450\\ \hline
3& 0& 1& 4& 4& 2& False&     6717\\ \hline
3& 0& 1& 4& 7& 1& False&     4618\\ \hline
3& 0& 1& 4& 1& 3& False&     4568\\ \hline
4& 1& 1& 4& 6& 4& False&     3934\\ \hline
4& 1& 1& 4& 1& 1& False&     3753\\ \hline
3& 0& 2& 5& 7& 3& False&     3294\\ \hline
3& 0& 1& 4& 2& 4& False&     2496\\ \hline
3& 0& 1& 4& 1& 1& False&     2418\\ \hline
3& 0& 1& 4& 2& 2& False&     1997\\ \hline
3& 0& 1& 4& 5& 1& False&     1951\\ \hline
4& 1& 1& 4& 4& 6& False&     1620\\ \hline
3& 0& 2& 5& 3& 7& False&     1278\\ \hline
3& 0& 2& 5& 3& 3& False&     1137\\ \hline
3& 0& 1& 4& 1& 7& False&     1052\\ \hline
3& 0& 1& 4& 1& 5& False&      904\\ \hline
3& 0& 2& 5& 5& 1& False&      834\\ \hline
4& 1& 1& 4& 4& 4& False&      578\\ \hline
4& 1& 1& 4& 2& 2& False&      571\\ \hline
5& 2& 1& 4& 7& 5& False&      555\\ \hline
4& 1& 1& 4& 5& 1& False&      530\\ \hline
4& 1& 1& 4& 9& 1& False&      493\\ \hline
4& 1& 2& 5& 7& 5& False&      493\\ \hline
4& 1& 1& 4& 3& 1& False&      336\\ \hline
5& 2& 1& 4& 5& 7& False&      334\\ \hline
4& 1& 1& 4& 1& 5& False&      287\\ \hline
3& 0& 2& 5& 1& 5& False&      267\\ \hline
3& 0& 2& 5& 1& 1& False&      244\\ \hline
4& 1& 2& 5& 5& 7& False&      240\\ \hline
5& 2& 1& 4& 1& 1& False&      219\\ \hline
4& 1& 2& 5& 1& 1& False&      209\\ \hline
4& 1& 1& 4& 6& 6& False&      176\\ \hline
5& 2& 1& 4& 5& 5& False&      162\\ \hline
4& 1& 2& 5& 5& 5& False&      120\\ \hline
4& 1& 1& 4& 1& 9& False&      108\\ \hline
4& 1& 1& 4& 7& 7& False&      104\\ \hline
3& 0& 2& 5& 6& 2& False&       80\\ \hline
4& 1& 2& 5& 6& 4& False&       78\\ \hline
5& 2& 1& 4& 6& 4& False&       67\\ \hline
4& 1& 2& 5& 3& 1& False&       59\\ \hline
3& 0& 2& 5& 2& 6& False&       55\\ \hline
5& 2& 1& 4& 3& 1& False&       52\\ \hline
3& 0& 2& 5& 2& 2& False&       39\\ \hline
4& 1& 2& 5& 4& 4& False&       26\\ \hline
5& 2& 1& 4& 4& 6& False&       22\\ \hline
4& 1& 2& 5& 4& 6& False&       20\\ \hline
4& 1& 1& 4& 7& 5& False&       16\\ \hline
4& 1& 2& 5& 1& 3& False&       11\\ \hline
3& 0& 3& 6& 3& 1& False&       11\\ \hline
3& 0& 3& 6& 3& 9& False&        3\\ \hline
\end{tabular}
\caption{\label{keyNumtableSt} \emph{Main characteristic quantum numbers of $\tilde{S}$-- satisfying all constraints (1)-(9) from Tables \ref{RandomSStats} and \ref{FertileSStats}}}
\end{table}

\begin{table}[!htb]
\scriptsize
\setlength{\tabcolsep}{5pt}
\renewcommand{\arraystretch}{0.85}
\centering
\begin{tabular}{|c|c|c|c|c|c|c|c|c|c|c|c|}
\hline
$n_{4L}$&$n_{\bar{4}L}$&$n_{4R}$&$n_{\bar{4}R}$&$n_h$&$n_6$&No Fermionic Exotics&Frequency\\\hline
3& 0& 1& 4& 5& 3& False&31059\\\hline
4& 1& 0& 3& 5& 3& False&31044\\\hline
3& 0& 1& 4& 3& 1& False&28885\\\hline
4& 1& 0& 3& 3& 1& False&28510\\\hline
3& 0& 0& 3& 2& 2& False&20833\\\hline
3& 0& 1& 4& 3& 5& False&18083\\\hline
4& 1& 0& 3& 3& 5& False&17269\\ \hline
4& 1& 1& 4& 3& 3& False&16309\\ \hline
4& 1& 1& 4& 1& 1& False&16230\\ \hline
4& 1& 1& 4& 5& 5& False&14956\\ \hline
3& 0& 1& 4& 3& 3& False&11068\\ \hline
4& 1& 0& 3& 3& 3& False&11033\\ \hline
3& 0& 1& 4& 1& 1& False&10871\\ \hline
3& 0& 0& 3& 3& 3& False&10614\\ \hline
4& 1& 0& 3& 1& 1& False&10379\\ \hline
3& 0& 1& 4& 1& 3& False&9237\\ \hline
3& 0& 0& 3& 1& 1& False&9045\\ \hline
4& 1& 0& 3& 1& 3& False&8610\\ \hline
4& 1& 0& 3& 4& 2& False&8317\\ \hline
3& 0& 1& 4& 4& 2& False&8098\\ \hline
4& 1& 1& 4& 2& 2& False&6928\\ \hline
4& 1& 1& 4& 5& 5& True&5841\\ \hline
4& 1& 0& 3& 7& 1& False&4658\\ \hline
3& 0& 1& 4& 7& 1& False&4512\\ \hline
4& 1& 1& 4& 4& 4& False&3616\\ \hline
4& 1& 0& 3& 2& 4& False&3111\\ \hline
3& 0& 1& 4& 2& 2& False&3043\\ \hline
3& 0& 1& 4& 2& 4& False&2897\\ \hline
4& 1& 1& 4& 6& 4& False&2742\\ \hline
4& 1& 0& 3& 5& 1& False&2508\\ \hline
4& 1& 0& 3& 2& 2& False&2439\\ \hline
3& 0& 1& 4& 5& 1& False&2379\\ \hline
3& 0& 0& 3& 5& 5& False&1992\\ \hline
4& 1& 1& 4& 3& 3& True&1960\\ \hline
4& 1& 0& 3& 5& 3& True&1944\\ \hline
3& 0& 0& 3& 4& 4& False&1918\\ \hline
5& 2& 1& 4& 3& 1& False&1692\\ \hline
4& 1& 2& 5& 3& 1& False&1656\\ \hline
3& 0& 1& 4& 5& 3& True&1564\\ \hline
5& 2& 0& 3& 7& 3& False&1382\\ \hline
3& 0& 2& 5& 7& 3& False&1366\\ \hline
3& 0& 0& 3& 3& 1& False&1227\\ \hline
3& 0& 1& 4& 1& 7& False&1188\\ \hline
3& 0& 1& 4& 1& 5& False&1164\\ \hline
4& 1& 0& 3& 1& 7& False&1090\\ \hline
4& 1& 0& 3& 1& 5& False&1084\\ \hline
4& 1& 1& 4& 3& 1& False&1053\\ \hline
4& 1& 2& 5& 1& 1& False&996\\ \hline
4& 1& 0& 3& 3& 5& True&974\\ \hline
4& 1& 1& 4& 6& 6& False&903\\ \hline
5& 2& 1& 4& 1& 1& False&890\\ \hline
4& 1& 1& 4& 1& 1& True&824\\ \hline
4& 1& 1& 4& 4& 6& False&778\\ \hline
3& 0& 2& 5& 3& 3& False&760\\ \hline
3& 0& 0& 3& 5& 1& False&757\\ \hline
5& 2& 0& 3& 3& 3& False&705\\ \hline
5& 2& 1& 4& 7& 5& False&690\\ \hline
3& 0& 2& 5& 5& 1& False&662\\ \hline
4& 1& 2& 5& 7& 5& False&660\\ \hline
5& 2& 0& 3& 5& 1& False&660\\ \hline
5& 2& 0& 3& 1& 1& False&660\\ \hline
3& 0& 0& 3& 4& 2& False&643\\ \hline
5& 2& 0& 3& 3& 7& False&620\\ \hline
3& 0& 1& 4& 3& 5& True&592\\ \hline
4& 1& 1& 4& 5& 1& False&580\\ \hline
3& 0& 2& 5& 3& 7& False&528\\ \hline
3& 0& 0& 3& 5& 5& True&512\\ \hline
4& 1& 2& 5& 6& 4& False&502\\ \hline
4& 1& 2& 5& 1& 3& False&499\\ \hline
4& 1& 0& 3& 3& 1& True&468\\ \hline
5& 2& 1& 4& 1& 3& False&456\\ \hline
3& 0& 1& 4& 3& 1& True&452\\ \hline
4& 1& 2& 5& 5& 7& False&440\\ \hline
3& 0& 2& 5& 1& 1& False&420\\ \hline
3& 0& 0& 3& 3& 3& True&415\\ \hline
\end{tabular}
\caption{\label{keyNumtableS} Part 1: \emph{Main characteristic quantum numbers of 411578 $S$--models satisfying all constraints (1)-(8) from Tables \ref{RandomSStats} and \ref{FertileSStats}}}
\end{table}

\begin{table}[!htb]
\scriptsize
\setlength{\tabcolsep}{5pt}
\renewcommand{\arraystretch}{0.85}
\centering
\begin{tabular}{|c|c|c|c|c|c|c|c|c|c|c|c|}
\hline
$n_{4L}$&$n_{\bar{4}L}$&$n_{4R}$&$n_{\bar{4}R}$&$n_h$&$n_6$&No Fermionic Exotics&Frequency\\\hline
5& 2& 1& 4& 5& 7& False&348\\ \hline
4& 1& 1& 4& 9& 1& False&328\\ \hline
4& 1& 1& 4& 7& 7& False&320\\ \hline
4& 1& 2& 5& 5& 5& False&314\\ \hline
5& 2& 1& 4& 6& 4& False&302\\ \hline
4& 1& 1& 4& 1& 5& False&252\\ \hline
3& 0& 2& 5& 1& 5& False&210\\ \hline
5& 2& 1& 4& 7& 5& True&208\\ \hline
4& 1& 2& 5& 7& 5& True&200\\ \hline
5& 2& 0& 3& 6& 2& False&188\\ \hline
5& 2& 1& 4& 3& 1& True&180\\ \hline
3& 0& 0& 3& 1& 3& False&177\\ \hline
4& 1& 2& 5& 4& 6& False&168\\ \hline
5& 2& 2& 5& 1& 1& False&168\\ \hline
5& 2& 0& 3& 1& 5& False&163\\ \hline
3& 0& 2& 5& 6& 2& False&148\\ \hline
4& 1& 0& 3& 1& 3& True&132\\ \hline
5& 2& 1& 4& 5& 7& True&120\\ \hline
3& 0& 1& 4& 1& 3& True&116\\ \hline
4& 1& 2& 5& 3& 1& True&112\\ \hline
4& 1& 2& 5& 5& 7& True&112\\ \hline
5& 2& 1& 4& 5& 5& False&108\\ \hline
4& 1& 1& 4& 1& 9& False&102\\ \hline
5& 2& 0& 3& 2& 6& False&98\\ \hline
3& 0& 0& 3& 7& 3& False&92\\ \hline
4& 1& 1& 4& 7& 5& False&86\\ \hline
5& 2& 1& 4& 1& 3& True&84\\ \hline
5& 2& 1& 4& 4& 6& False&80\\ \hline
3& 0& 0& 3& 7& 7& True&80\\ \hline
5& 2& 2& 5& 2& 2& False&72\\ \hline
3& 0& 1& 4& 4& 2& True&72\\ \hline
3& 0& 3& 6& 3& 1& False&72\\ \hline
4& 1& 0& 3& 4& 2& True&72\\ \hline
3& 0& 0& 3& 5& 3& False&71\\ \hline
5& 2& 0& 3& 2& 2& False&68\\ \hline
6& 3& 0& 3& 3& 1& False&62\\ \hline
3& 0& 0& 3& 6& 6& False&60\\ \hline
3& 0& 0& 3& 6& 2& False&58\\ \hline
3& 0& 2& 5& 2& 2& False&54\\ \hline
6& 3& 0& 3& 1& 1& False&48\\ \hline
4& 1& 1& 4& 2& 2& True&48\\ \hline
3& 0& 0& 3& 2& 4& False&45\\ \hline
3& 0& 3& 6& 1& 1& False&44\\ \hline
4& 1& 2& 5& 4& 4& False&36\\ \hline
3& 0& 0& 3& 1& 5& False&33\\ \hline
3& 0& 0& 3& 7& 7& False&24\\ \hline
3& 0& 0& 3& 9& 1& False&24\\ \hline
6& 3& 0& 3& 2& 8& False&24\\ \hline
3& 0& 3& 6& 3& 9& False&20\\ \hline
4& 1& 2& 5& 1& 3& True&16\\ \hline
6& 3& 0& 3& 3& 9& False&14\\ \hline
3& 0& 2& 5& 2& 6& False&12\\ \hline
6& 3& 0& 3& 1& 3& False&10\\ \hline
3& 0& 3& 6& 1& 3& False&10\\ \hline
3& 0& 3& 6& 1& 9& False&10\\ \hline
3& 0& 1& 4& 2& 4& True&8\\ \hline
4& 1& 2& 5& 6& 4& True&8\\ \hline
5& 2& 1& 4& 6& 4& True&8\\ \hline
4& 1& 0& 3& 2& 4& True&8\\ \hline
5& 2& 1& 4& 4& 4& False&4\\ \hline
3& 0& 0& 3& 1& 1& True&4\\ \hline
6& 3& 0& 3& 1& 11& False&4\\ \hline
3& 0& 0& 3& 3& 5& False&3\\ \hline
6& 3& 0& 3& 1& 9& False&2\\ \hline
3& 0& 0& 3& 11& 3& False&2\\ \hline
3& 0& 0& 3& 4& 4& True&1\\ \hline
3& 0& 0& 3& 2& 2& True&1\\ \hline
\end{tabular}
\caption{\label{keyNumtableS2} Part 2: \emph{Main characteristic quantum numbers of 411578 $S$--models satisfying all constraints (1)-(8) from Tables \ref{RandomSStats} and \ref{FertileSStats}}}
\end{table}
\normalsize
\bibliographystyle{unsrt}

\end{document}